\documentclass[journal]{IEEEtran}
\usepackage[T1]{fontenc}
\usepackage{amsmath,amssymb,amsthm,mathtools,bm}
\usepackage{mathrsfs}
\usepackage{graphicx,booktabs,array}
\usepackage[caption=false,font=footnotesize]{subfig}
\usepackage{algorithm,algorithmic}
\usepackage{placeins}
\usepackage{cite}
\usepackage{xurl}
\usepackage[hidelinks]{hyperref}
\allowdisplaybreaks[1]
\graphicspath{{figure/}}

\DeclareMathOperator{\tr}{tr}
\DeclareMathOperator{\diag}{diag}

\DeclareMathOperator{\vecop}{vec}

\newcommand{\E}{\mathbb{E}}

\newcommand{\C}{\mathbb{C}}
\newcommand{\I}{\mathbf{I}}
\newcommand{\zero}{\mathbf{0}}
\newcommand{\T}{\mathsf{T}}
\newcommand{\Hh}{\mathsf{H}}

\newtheorem{proposition}{Proposition}

\theoremstyle{definition}
\newtheorem{remark}{Remark}

\begin{document}
\bstctlcite{jdcc:bstcontrol}

\title{Joint Design of Communication and Control in Center-Fed Pinching-Antenna Systems}

\author{Xu Gan,~\IEEEmembership{Member, IEEE}, Yuanwei Liu,~\IEEEmembership{Fellow,~IEEE}, and Hyundong Shin,~\IEEEmembership{Fellow,~IEEE}
		\thanks{Xu Gan and Yuanwei Liu are with the Department of Electrical and Computer Engineering, The University of Hong Kong, Hong Kong (e-mail: \{eee.ganxu, yuanwei\}@hku.hk).}
		\thanks{Hyundong Shin is with the Department of Electronic Engineering, Kyung Hee University, Yongin 17104, South Korea (e-mail: hshin@khu.ac.kr).}
    \vspace{-5mm}
	}
  
\maketitle

\begin{abstract}
An analytical and optimization framework for the joint design of communication and control (JDCC) in a center-fed pinching-antenna system (C-PASS) is proposed. Within this framework, the mathematical coupling between data throughput and closed-loop control cost is characterized for JDCC.
The analysis captures the mutual interference between
concurrently transmitted data and control streams during
uplink state uploading and downlink command delivery.
Based on this coupling, an achievable communication–control performance region is established for C-PASS. 
Accordingly, a throughput maximization problem is formulated to jointly optimize digital and pinching beamforming subject to control-cost constraints.
To address this problem, an alternating optimization algorithm is developed based on constraint reformulation and resource decomposition.
The proposed framework reveals how control requirements shape achievable data throughput, guiding the joint beamforming design.
Numerical results demonstrate that 1) the proposed C-PASS
design achieves 110.2\% higher frame-average throughput than
conventional PASS, and 2) the proposed concurrent
data-control transmission protocol increases frame-average
throughput by up to 66.7\% over separated transmission.
\end{abstract}


\section{Introduction}\label{sec:introduction}
\IEEEPARstart{I}{t} is envisioned that sixth-generation (6G) networks will support
emerging applications, such as intelligent manufacturing, networked robotics,
and remote operation, that require both high-throughput data transmission
and reliable closed-loop control~\cite{itu2023imt2030}. In remote operation,
for example, visual data support situational awareness, while timely state
updates and feedback commands enable closed-loop regulation of the remote
device~\cite{itu2023imt2030,park2018wireless}. However, supporting these two
functions over a common wireless infrastructure entails shared resource
constraints and potential interference between data and control
transmissions~\cite{echevarria2023robust}. In particular, resource
allocation and interference affect both data rates and packet loss in
control exchanges, which in turn influence data
throughput and closed-loop control performance,
respectively~\cite{liu2021latency}. This coupling motivates the joint
design of communication and control~\cite{gan2026jdcc}, which enables more efficient use
of shared wireless resources through coordinated data and control
transmissions.

\subsection{Related Work}
To assess progress toward this joint design and identify remaining
challenges, we review communication-control performance modeling
and resource-allocation methods relevant to data and control coexistence.

\subsubsection{Communication-Control Performance Modeling}
To model the coupling between communication constraints and control
performance, Chen \emph{et al.}~\cite{chen2025formation} characterized
guaranteed communication regions under control-system uncertainty
and derived data-rate limits for prescribed formation accuracy.
For unreliable bidirectional links, Cao
\emph{et al.}~\cite{cao2025goal} modeled the evolution of estimation
error and plant-state covariance in terms of update age and
successful packet delivery, using plant-state violation probability
as a control-performance metric. To capture uplink--downlink
dependence, Meng \emph{et al.}~\cite{meng2026closedloop} modeled
closed-loop delay and packet loss using effective-capacity theory
and derived a sufficient convergence condition relating bandwidth,
quantization, and control parameters. Under shared data and control
resources, our earlier work~\cite{gan2026jdcc} modeled state and
command reconstruction through rate-distortion relationships and
characterized the communication-delay--control-variance Pareto
boundary and joint outage probability. Besides, Jiang
\emph{et al.}~\cite{jiang2026jcc} modeled uplink Kalman estimation
and downlink actuation distortion within a closed-loop system and
derived finite- and infinite-horizon linear quadratic Gaussian (LQG)
costs under communication-control interference.

\subsubsection{Resource Allocation for Communication and Control}
To allocate wireless resources according to control-task requirements,
Han \emph{et al.}~\cite{han2024identification} jointly designed power
and channel allocation for throughput maximization or power
minimization with finite-time system-identification accuracy
constraints. Feng \emph{et al.}~\cite{feng2024goal} allocated bandwidth
to maximize task-related information utility, combining utility
decomposition with greedy selection. For coupled control loops,
Pang \emph{et al.}~\cite{pang2025codesign} used deep reinforcement
learning to jointly optimize transmission scheduling and control
inputs over Markov block-fading channels. Fang
\emph{et al.}~\cite{fang2025closedloop} jointly allocated uplink,
downlink, and computing resources within and across control loops to
minimize total linear quadratic regulator (LQR) cost, combining
closed-form intra-loop allocation with iterative inter-loop
optimization. Under state-dependent Markov fading, Wang
\emph{et al.}~\cite{wang2026markovcodesign} co-designed multi-frequency
transmission and mobile-agent policies to minimize average joint
communication-control cost subject to mean-square stability,
using a bilinear matrix inequality formulation.

The performance-modeling studies reviewed above quantify the relationships
between communication constraints and closed-loop control performance,
while the resource-allocation studies develop communication
resource-allocation schemes guided by control-task requirements.
However, the joint communication-control performance achievable with these
schemes remains limited by the propagation conditions of the data and
control links. To improve these propagation conditions, antenna reconfiguration
and resource allocation for these two functions can be jointly optimized
to strengthen desired links and mitigate mutual interference.
As a promising reconfigurable antenna architecture, the pinching-antenna system
(PASS) provides additional spatial degrees of freedom for channel shaping
and interference management~\cite{liu2026pass}. Specifically, pinching
antennas (PAs) couple guided signals from dielectric waveguides into free
space, and repositioning these antennas adjusts the radiating locations
and hence the propagation channels~\cite{xu2025attenuation}. We next review
resource allocation in PASS and the potential of center-fed implementations
for practical deployment.

\subsubsection{PASS and C-PASS Resource Allocation}
By relocating radiating points along waveguides, PASS can shorten
free-space transmission distances and adapt propagation channels to user
locations. Exploiting this spatial flexibility, Shan
\emph{et al.}~\cite{shan2026multicast} maximized the worst-user rate
through joint transmit and pinching beamforming, combining
majorization-minimization updates with sequential PA optimization.
To manage inter-group interference, the study in~\cite{shan2026multigroup}
jointly optimized PA placement and power allocation under non-orthogonal
multiple access and developed time-division protocols with group-specific
or shared PA configurations and optimized time-power allocation.
For lossy waveguides, Xu \emph{et al.}~\cite{xu2025attenuation}
incorporated attenuation into the channel model and jointly optimized
PA placement and transmit beamforming for downlink sum-rate maximization.
To exploit guided propagation as an additional design dimension, Gan
and Liu~\cite{gan2025c} proposed the center-fed pinching-antenna system
(C-PASS), where signals injected at interior feed ports propagate in
both waveguide directions to support spatial multiplexing. The study
in~\cite{gan2026center} introduced controllable feed-power splitting and
switching protocols and jointly optimized transmit and pinching
beamforming for sum-rate maximization. The generalized C-PASS
framework in~\cite{gan2026model} further characterized the spatial
degrees of freedom and power scaling with multiple feed ports and
jointly optimized digital precoding, feed-power splitting, PA positions,
and radiation coefficients. On the uplink, Yu
\emph{et al.}~\cite{yu2026environment} characterized feed-point spacing
for stable signal separation and derived Ziv--Zakai bounds on
environment-reconstruction error.

\subsection{Motivation and Contributions}
The studies reviewed above have quantified the dependence of closed-loop control
performance on communication quality and developed wireless resource-allocation
methods guided by control-task requirements, while PASS/C-PASS research
has optimized antenna configuration and signal processing for communication
and sensing. However, a fundamental performance analysis of
communication--control tradeoffs in C-PASS and a systematic joint design
framework remain lacking. To fill this gap, this article establishes an
achievable performance region for communication and control coexistence and
exploits C-PASS reconfiguration to increase data throughput while
maintaining the prescribed closed-loop control performance.
The main contributions are as follows.
\begin{itemize}
\item We develop a unified performance model for communication and
control coexistence among multiple data users and control devices
sharing a reciprocal C-PASS configuration. The model captures mutual
interference between data and control transmissions and accounts for
communication delay and packet loss. We characterize data throughput
averaged over each frame and derive the steady-state quadratic control
cost in terms of round-trip delivery reliability.
\item Building on this performance model, we establish an inner
communication--control performance region that characterizes the data
throughput achievable under prescribed control cost limits. To construct
this region, we translate the control cost limits into sufficient uplink
and downlink signal-to-interference-plus-noise ratio (SINR) constraints
using a finite-blocklength normal approximation. These constraints limit
the interference from concurrent data transmissions to protect control
performance.
\item Using the performance model and the established inner performance
region, we formulate a joint optimization problem to maximize data throughput
averaged over each frame subject to the prescribed control cost limit
of each device. To solve this coupled nonconvex problem, we develop an
alternating optimization algorithm that integrates sequential quadratic
programming for joint reliability and transceiver optimization with
quasi-Newton updates for C-PASS reconfiguration. The proposed algorithm thus
incorporates closed-loop control requirements into C-PASS reconfiguration
and exploits the joint adaptation of propagation channels and transmission
resources to improve data throughput.
\item Numerical results validate the control-cost analysis and quantify
the frame-average throughput gains of the proposed design. Under the
considered configurations and protocols, C-PASS achieves 110.2\% higher
frame-average throughput than conventional PASS. Concurrent data and
control transmission increases frame-average throughput by up to 66.7\%
relative to separated transmission.
\end{itemize}

\section{System and Transmission Models}\label{sec:system}
This section establishes the system and transmission models for
C-PASS-assisted communication and closed-loop control. We first specify
the system topology, C-PASS configuration, and reciprocal channels.
We then present the protocol for concurrent state and data uploading,
concurrent command and data delivery, and data-only transmission.
Based on these models and protocol, we derive the uplink and downlink
signals and SINRs for the subsequent performance analysis.

\subsection{System Model}
\begin{figure}[!t]
\centering
\includegraphics[width=0.9\columnwidth]{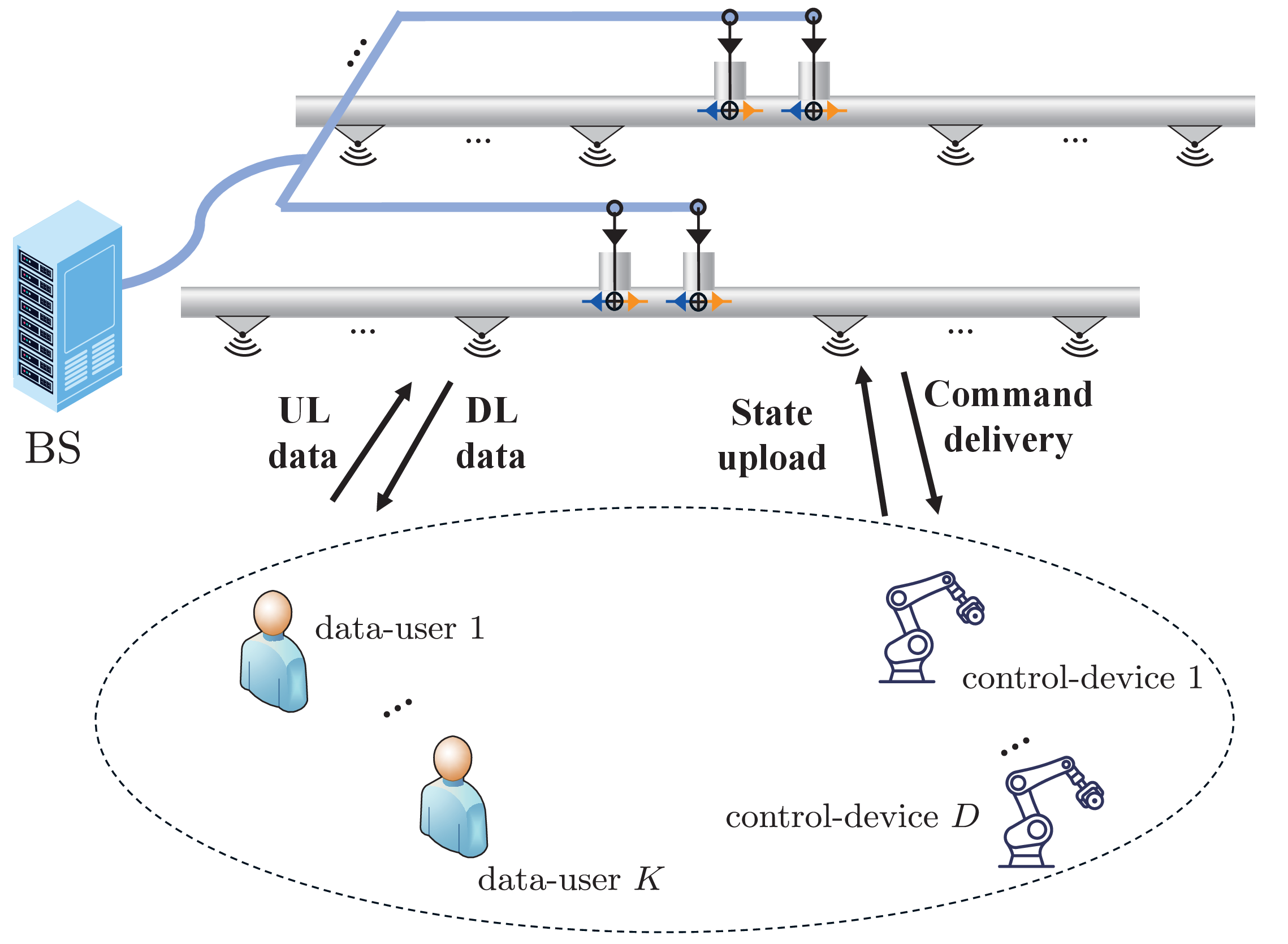}
\caption{Illustration of the C-PASS-assisted communication and control system.}
\label{fig:system-model}
\end{figure}

As illustrated in Fig.~\ref{fig:system-model}, a base station (BS)
feeds radio-frequency (RF) signals into the dielectric waveguides of
a C-PASS through centrally located feed ports. The guided signals are then
coupled into free space by PAs distributed along the waveguides at a
common height \(H\) above the ground to serve \(D\) single-antenna control
devices and \(K\) single-antenna data users, indexed by
\(\mathcal D=\{1,\ldots,D\}\) and \(\mathcal K=\{1,\ldots,K\}\),
respectively. For notational simplicity, the control and data stream sets are
\(\mathcal R_{\rm c}=\mathcal D\) and
\(\mathcal R_{\rm d}=\{D+j:j\in\mathcal K\}\), with
\(\mathcal R=\mathcal R_{\rm c}\cup\mathcal R_{\rm d}\). Terminal \(r\in\mathcal R\) has coordinates
\((x_r^{\rm t},y_r^{\rm t},z_r^{\rm t})\) relative
to ground \(z=0\), with horizontal position in the rectangular service
area \([0,L_x]\times[0,L_y]\).

Within this C-PASS architecture, the BS jointly combines the signals
received at the feed ports during uplink (UL) transmission to detect
the data and control streams. For downlink (DL) transmission, it jointly
precodes data and control streams across the same ports for delivery
to the data users and control devices. Through these bidirectional transmissions, the BS provides
data services and connects the control devices to an edge controller
that computes feedback commands for their plants and local actuators.
To characterize the desired signal powers and mutual interference,
the following subsections establish the reciprocal C-PASS channel model,
specify the transmission protocol, and derive the corresponding UL and
DL signal models.

\subsection{C-PASS Configuration and Channel Model}
This subsection provides the expressions for the reciprocal C-PASS channels in terms of
the physical configuration. 
Specifically, the C-PASS comprises \(N\) movable PAs on \(L\) waveguides, with two fixed
central feed ports per waveguide. Each of the \(M=2L\) ports is connected
to a calibrated RF chain. The physical configuration
\(\boldsymbol\varphi\) collects the PA positions
\(\mathbf q=[q_1,\ldots,q_N]^{\T}\), forward and backward feed-coupling
amplitudes \(\boldsymbol\rho^{\rm F}\) and \(\boldsymbol\rho^{\rm B}\),
and PA extraction fractions \(\boldsymbol\zeta\). For this configuration, the equivalent channel vector from
the \(M\) feed ports to the \(r\)-th terminal is given by
\begin{equation}\label{eq:cpass-channel}
 \mathbf f_r(\boldsymbol\varphi)
 =\mathbf E(\boldsymbol\varphi)^{\T}\mathbf h_r^{\rm w}(\mathbf q),
\end{equation}
where \(\mathbf E(\boldsymbol\varphi)\in\C^{N\times M}\) is the normalized
transfer matrix from the \(M\) feed ports to the \(N\) PAs, and
\(\mathbf h_r^{\rm w}(\mathbf q)
=[h_{1,r}^{\rm w}(q_1),\ldots,h_{N,r}^{\rm w}(q_N)]^{\T}\)
is the wireless channel vector from the PAs to the \(r\)-th terminal.

$\bullet$ \textit{Feed-to-PA transfer.}
To characterize \(\mathbf E(\boldsymbol\varphi)\)
in~\eqref{eq:cpass-channel}, let \(\mathcal N_l\) denote the PA index
set of the \(l\)-th waveguide. The \(m\)-th feed port is fixed at
coordinate \(d_m^{\rm feed}\), while the \(n\)-th PA has adjustable
coordinate \(q_n\). The forward and backward PA sets
\(\mathcal N_m^{\rm F}\) and \(\mathcal N_m^{\rm B}\) contain the PAs
on the same waveguide to the right and left of the \(m\)-th feed port,
respectively. Placement and spacing constraints preserve these sets
and the PA order during reconfiguration. Thus, the transfer coefficient is
\begin{equation}\label{eq:dual-feed-transfer}
 E_{nm}(\boldsymbol\varphi)=
 \rho_m^{\chi}\sqrt{\xi_{m,n}^{\chi}(\boldsymbol\zeta)}
 g_{m,n}^{\rm wg},\qquad n\in\mathcal N_m^{\chi},
\end{equation}
where \(\chi\in\{\mathrm F,\mathrm B\}\) denotes the propagation
direction and \(\rho_m^{\chi}\geq0\) is the feed-coupling amplitude,
with directional power fraction \((\rho_m^{\chi})^2\). The \(n\)-th PA
extracts a fraction \(\zeta_n\) of the power remaining after the
intervening PAs, giving~\cite{gan2026model}
\begin{equation}\label{eq:radiation-simplex}
 \xi_{m,n}^{\chi}(\boldsymbol\zeta)
 =\zeta_n\prod_{v\in\mathcal P_{m,n}}(1-\zeta_v),
 \qquad n\in\mathcal N_m^{\chi},
\end{equation}
where \(\mathcal P_{m,n}\) contains the PAs between the \(m\)-th feed
port and the \(n\)-th PA along direction \(\chi\). The extraction
fraction \(\zeta_n\) is common to all feed ports and both propagation
directions.
The remaining guided power is terminated or dissipated. Propagation over
the feed--PA distance contributes the factor
\(g_{m,n}^{\rm wg}=\exp(-c_{\rm wg}|q_n-d_m^{\rm feed}|)\),
where \(c_{\rm wg}=\alpha_{\rm wg}/2+\mathrm j k_{\rm wg}\) uses the
power-attenuation coefficient \(\alpha_{\rm wg}\) and guided wavenumber
\(k_{\rm wg}=2\pi/\lambda_{\rm wg}\). The guided wavelength is
\(\lambda_{\rm wg}=c_0/(n_{\rm eff}f_c)\), where \(c_0\) is the speed of
light, \(n_{\rm eff}\) is the effective refractive index, and \(f_c\) is
the carrier frequency.

$\bullet$ \textit{PA-to-terminal channel.}
To characterize \(\mathbf h_r^{\rm w}(\mathbf q)\)
in~\eqref{eq:cpass-channel}, we assume a single line-of-sight (LoS) path
from the \(n\)-th PA to the \(r\)-th terminal. The free-space channel
coefficient is~\cite{gan2026model}
\begin{equation}\label{eq:los-wireless-channel}
 h_{n,r}^{\rm w}(q_n)=\frac{\lambda_0}{4\pi d_{n,r}}
 \exp\!\left(-\mathrm j\frac{2\pi d_{n,r}}{\lambda_0}\right),
\end{equation}
where \(\lambda_0=c_0/f_c\) and \(y_l\) is the transverse coordinate of
the \(l\)-th waveguide. For \(n\in\mathcal N_l\), the propagation distance is
\(d_{n,r}=\sqrt{(q_n-x_r^{\rm t})^2+(y_l-y_r^{\rm t})^2+(H-z_r^{\rm t})^2}\).

$\bullet$ \textit{Hardware configuration constraints.}
The physical configuration \(\boldsymbol\varphi\) in the channel model
is subject to PA placement intervals and minimum spacing,
feed-coupling power limits, PA extraction bounds, and passivity
of the feed-to-PA transfer.
The resulting set of admissible C-PASS configurations is
\begin{equation}\label{eq:hardware-feasible-set}
\begin{aligned}
 \mathcal X_{\rm C}^{\rm hw}=\bigl\{&\boldsymbol\varphi:\;
 q_n\in[q_n^{\min},q_n^{\max}],\quad
 \zeta_n\in[\zeta_{\min},\zeta_{\max}],\\
 &q_{n+1}-q_n\geq d_l^{\min},\quad n,n+1\in\mathcal N_l,\\
 &\rho_m^{\chi}\geq0,\;
 (\rho_m^{\rm F})^2+(\rho_m^{\rm B})^2\leq1,\;
 \mathbf E^{\Hh}\mathbf E\preceq\I_M\bigr\}.
\end{aligned}
\end{equation}
Here, \([q_n^{\min},q_n^{\max}]\subseteq[0,\ell_l]\) is the placement
interval of the \(n\)-th PA on the \(l\)-th waveguide of length \(\ell_l\).
PAs are indexed consecutively in spatial order, with minimum spacing
\(d_l^{\min}\). The feed constraint limits the combined forward and
backward power fractions to one, while
\(0<\zeta_{\min}\leq\zeta_{\max}<1\) bounds PA extraction.
The coupling and extraction bounds limit individual power fractions.
Since signals from different feed ports can add coherently at the shared
PAs, the additional constraint
\(\mathbf E^{\Hh}\mathbf E\preceq\I_M\) enforces passivity of the
complete network by ensuring that the total PA output power never
exceeds the total feed input power.
To characterize UL reception through the same C-PASS configuration,
we use electromagnetic reciprocity in
Proposition~\ref{prop:reciprocal-feeding} to relate the terminal-to-feed
channel to the feed-to-terminal model
in~\eqref{eq:cpass-channel}.

\begin{proposition}[Reciprocity]
\label{prop:reciprocal-feeding}
For any fixed \(\boldsymbol\varphi\in\mathcal X_{\rm C}^{\rm hw}\)
of the dielectric C-PASS considered here,
let \(\mathbf h_r^{\rm u}\in\C^{M\times1}\) denote the UL channel
from terminal \(r\in\mathcal R\) to the \(M\) BS feed ports, and let
\(\mathbf h_r^{\rm d}\in\C^{M\times1}\) denote the DL channel from
the \(M\) BS feed ports to terminal \(r\). These channels are reciprocal
and are respectively expressed as
\[
\begin{aligned}
 \mathbf h_r^{\rm u}
 &=\mathbf E^{\T}\mathbf h_r^{\rm w}=\mathbf f_r,\\
 \mathbf h_r^{\rm d}
 &=\mathbf E^{\Hh}(\mathbf h_r^{\rm w})^*=\mathbf f_r^*.
\end{aligned}
\]
\end{proposition}
\begin{proof}
Please refer to Appendix~\ref{app:reciprocity-proof}.
\end{proof}
The effective UL and DL channel gains are jointly determined by the
C-PASS configuration and the transceiver design.

\subsection{Communication--Control Transmission Protocol}
This subsection presents the communication--control transmission
protocol, which coordinates round-trip control updates and data access
within each frame. Specifically, in each frame of duration \(T\),
the devices first upload their sampled states over the UL, the edge
controller then computes feedback commands, and the BS delivers these
commands over the DL for actuation. Under this schedule, data traffic
shares the UL and DL control exchanges and uses the computation interval
and remaining frame time for data-only access.
We next specify the procedures for concurrent UL transmission,
concurrent DL transmission, and data-only access.

\subsubsection{Concurrent Control State and Data Uploading}
The BS receives state packets from the control devices concurrently
with user data, and the edge controller uses the decoded states to
compute feedback commands.
At \(t_k^{\rm s}=kT\), all \(D\) devices sample their states and upload
their packets over the UL interval of duration \(\tau^{\rm u}\).
The BS decodes the received streams using digital combiners.
The edge controller then computes feedback commands from the decoded
device states during \(\tau^{\rm e}\).
The control-packet sizes \(L_i^a\), directional bandwidths
\(B^a\), and interval durations are prescribed, where
\(a\in\mathcal A=\{\mathrm u,\mathrm d\}\) denotes UL or DL.

\subsubsection{Concurrent Control Command and Data Delivery}
The BS transmits the computed commands concurrently with DL data during
\(\tau^{\rm d}\), using a beamformer for each stream.
The command packet for control device \(i\) contains \(L_i^{\rm d}\)
information bits. All command and data streams share the DL bandwidth
\(B^{\rm d}\), so each terminal receives a superposition of its
intended stream and the other transmitted streams. The control devices
and data users decode their respective streams, treating the remaining
streams as interference.

\subsubsection{Data-Only Transmission}
Apart from the two control exchanges, the remaining frame intervals
are allocated exclusively to UL and DL data transmission.
A prescribed time-division duplex (TDD) schedule allocates these intervals
between UL and DL, yielding aggregate durations \(\tau_0^{\rm u}\) and
\(\tau_0^{\rm d}\) that satisfy
\(\tau_0^{\rm u}+\tau_0^{\rm d}=T-\tau^{\rm u}-\tau^{\rm d}\).
The BS uses digital combiners for UL reception and beamformers for DL
data delivery. These digital resources are designed separately from
those used in concurrent transmission, while the C-PASS configuration
remains unchanged.

\subsection{Uplink and Downlink Signal Models}
Under the transmission protocol described above, we derive the received
signals and corresponding SINRs for concurrent state and data uploading,
concurrent command and data delivery, and data-only transmission as follows.
For notational convenience, the SINRs of control device
\(i\in\mathcal D\) and data user \(j\in\mathcal K\) in either
direction \(a\in\mathcal A\) are denoted by
\(\gamma_i^{\mathrm{ctrl},a}=\gamma_i^a\) and
\(\gamma_j^{\mathrm{com},a}=\gamma_{D+j}^a\), respectively.

\subsubsection{Uplink Control State and Data Uploading}
During \(\tau^{\rm u}\), the control devices upload state packets
concurrently with the data users' streams, giving the BS received signal
\begin{equation}\label{eq:uplink-received-signal}
 \mathbf y^{\rm u}=\sum_{r\in\mathcal R}\sqrt{P_r}
 \mathbf h_r^{\rm u}s_r^{\rm u}+\mathbf n^{\rm u},
\end{equation}
where \(s_r^{\rm u}\) is a unit-variance symbol carrying state
information for \(r\in\mathcal R_{\rm c}\) or user data for
\(r\in\mathcal R_{\rm d}\), \(P_r\) is terminal \(r\)'s transmit power satisfying
\(0\leq P_r\leq P_{r,\max}^{\rm u}\), and
\(\mathbf n^{\rm u}\sim\mathcal{CN}(\zero,\sigma_{\rm u}^2\I_M)\) is the
BS port-noise vector, with \(\sigma_{\rm u}^2=N_0B^{\rm u}F_{\rm BS}\).
Here, \(N_0\) is the noise power spectral density and \(F_{\rm BS}\)
is the BS receiver noise factor.
The BS detects stream \(r\) from
\(\mathbf v_r^{\Hh}\mathbf y^{\rm u}\) using a combiner satisfying
\(\|\mathbf v_r\|=1\), yielding the SINR
\begin{equation}\label{eq:control-uplink-sinr}
 \gamma_r^{\rm u}=
 \frac{P_r|\mathbf v_r^{\Hh}\mathbf h_r^{\rm u}|^2}
 {\sum_{\ell\in\mathcal R\setminus\{r\}}P_\ell
 |\mathbf v_r^{\Hh}\mathbf h_\ell^{\rm u}|^2+\sigma_{\rm u}^2}.
\end{equation}
The denominator includes interference from all other control and
communication data streams.

\subsubsection{Downlink Control Command and Data Delivery}
During \(\tau^{\rm d}\), the BS concurrently delivers commands to the
control devices and data to the users, with transmit vector
\begin{equation}\label{eq:downlink-transmitted-signal}
 \mathbf x^{\rm d}=\sum_{r\in\mathcal R}\mathbf w_r s_r^{\rm d},
\end{equation}
where \(s_r^{\rm d}\) is a unit-variance symbol carrying a control
command for \(r\in\mathcal R_{\rm c}\) or user data for
\(r\in\mathcal R_{\rm d}\), and \(\mathbf w_r\in\C^M\) is its
beamformer. All control streams remain active in every frame. The
beamformers satisfy the BS power constraint
\(\sum_{r\in\mathcal R}\|\mathbf w_r\|^2\leq P_{\max}^{\rm d}\).
The received signal at terminal \(r\) is
\(y_r^{\rm d}=(\mathbf h_r^{\rm d})^{\Hh}\mathbf x^{\rm d}+n_r^{\rm d}\),
where \(n_r^{\rm d}\sim\mathcal{CN}(0,\sigma_r^2)\) and
\(\sigma_r^2=N_0B^{\rm d}F_r>0\), with \(F_r\) denoting the receiver
noise factor at terminal \(r\). Terminal \(r\) decodes its intended
stream with SINR
\begin{equation}\label{eq:control-downlink-sinr}
 \gamma_r^{\rm d}=
 \frac{|(\mathbf h_r^{\rm d})^{\Hh}\mathbf w_r|^2}
 {\sum_{\ell\in\mathcal R\setminus\{r\}}
  |(\mathbf h_r^{\rm d})^{\Hh}\mathbf w_\ell|^2+\sigma_r^2}.
\end{equation}
The denominator includes interference from all other control and
communication data streams.

\subsubsection{Data-Only Transmission}
During \(\tau_0^{\rm u}\) and \(\tau_0^{\rm d}\), the users upload and
receive data, respectively, under the same C-PASS configuration.
The corresponding UL and DL SINRs of user \(j\in\mathcal K\) are
\begin{equation}\label{eq:data-only-uplink-sinr}
 \gamma_j^{\mathrm{com,u},0}=
 \frac{P_{D+j}^0|(\mathbf v_{D+j}^0)^{\Hh}\mathbf h_{D+j}^{\rm u}|^2}
 {\sum_{\ell\in\mathcal R_{\rm d}\setminus\{D+j\}}P_\ell^0
 |(\mathbf v_{D+j}^0)^{\Hh}\mathbf h_\ell^{\rm u}|^2+\sigma_{\rm u}^2},
\end{equation}
\begin{equation}\label{eq:data-only-downlink-sinr}
 \gamma_j^{\mathrm{com,d},0}=
 \frac{|(\mathbf h_{D+j}^{\rm d})^{\Hh}\mathbf w_{D+j}^0|^2}
 {\sum_{\ell\in\mathcal R_{\rm d}\setminus\{D+j\}}
 |(\mathbf h_{D+j}^{\rm d})^{\Hh}\mathbf w_\ell^0|^2+\sigma_{D+j}^2},
\end{equation}
where \(P_r^0\), \(\mathbf v_r^0\), and \(\mathbf w_r^0\) are the
data-only UL power, receive combiner, and DL beamformer for stream
\(r\in\mathcal R_{\rm d}\), respectively. Only other data streams
contribute interference. These resources satisfy
\(0\leq P_{D+j}^0\leq P_{D+j,\max}^{\rm u}\),
\(\|\mathbf v_{D+j}^0\|=1\), and
\(\sum_{j\in\mathcal K}\|\mathbf w_{D+j}^0\|^2\leq P_{\max}^{\rm d}\).
The UL powers, UL combiners, and DL beamformers across all transmission
intervals are collected in \(\boldsymbol\Theta\), with feasible set
\(\mathcal T\) defined by the above constraints.

This section has established the C-PASS system model and
communication--control transmission protocol and derived the UL and DL
SINRs for concurrent and data-only transmission. Based on the above
system and transmission models, the subsequent section characterizes
the frame-average data throughput supported under closed-loop
control-cost limits.

\section{Performance Metrics and Achievable Region}\label{sec:performance}
This section characterizes the frame-average data throughput supported
by the shared C-PASS under prescribed control-cost limits. We first
define the communication and control performance metrics and derive
the control-cost--reliability relation under delayed actuation and
command holding. We then convert the cost limits into sufficient
directional SINR constraints and combine them with the throughput
constraint to construct an achievable inner region.

\subsection{Communication and Control Performance Metrics}
This subsection defines the communication and control performance
metrics based on the transmission protocol and signal models.
Specifically, we use frame-average weighted data throughput and control
cost to quantify communication and control performance, respectively.

\subsubsection{Frame-Average Data Throughput}
This metric measures the aggregate user-data rate across the concurrent
and data-only intervals of each frame. With Gaussian signaling and
interference treated as noise, averaging the weighted UL and DL data
rates over the frame yields
\begin{align}
 C(\boldsymbol\varphi,\boldsymbol\Theta)
 =&\,\frac{1}{T}\sum_{a\in\mathcal A}\biggl[
 \tau_0^a B^a\sum_{j\in\mathcal K}\omega_j
 \log_2(1+\gamma_j^{\mathrm{com},a,0})\nonumber\\
 &+\tau^a B^a\sum_{j\in\mathcal K}\omega_j
 \log_2(1+\gamma_j^{\mathrm{com},a})\biggr],
 \label{eq:communication-utility}
\end{align}
where \(\omega_j>0\) specifies user \(j\)'s relative service priority.

\subsubsection{Long-Term Quadratic Control Cost}
This metric evaluates control performance over repeated control cycles.
Each cycle contributes a cost associated with state deviations and
the applied control input.
For device \(i\), the cost is expressed as~\cite{anderson2007optimal,schenato2007foundations}
\begin{equation}\label{eq:control-cost-metric}
 J_i
 =\limsup_{N_{\rm c}\rightarrow\infty}\frac{1}{N_{\rm c}}
 \sum_{k=0}^{N_{\rm c}-1}\E\!\left[
 \mathbf x_{i,k}^{\T}\mathbf Q_i\mathbf x_{i,k}
 +\mathbf u_{i,k}^{\T}\mathbf R_i\mathbf u_{i,k}\right],
\end{equation}
where \(N_{\rm c}\) is the number of control cycles,
\(\mathbf x_{i,k}\) is device \(i\)'s sampled state conveyed by its UL
control packet, and \(\mathbf u_{i,k}\) is the control input actually
applied by its actuator at the \(k\)-th actuation deadline.
The first term, \(\mathbf x_{i,k}^{\T}\mathbf Q_i\mathbf x_{i,k}\),
measures the device state's deviation from its desired operating point,
while the second term, \(\mathbf u_{i,k}^{\T}\mathbf R_i\mathbf u_{i,k}\),
measures the control effort exerted by the actuator.
The prescribed matrices \(\mathbf Q_i\succeq\zero\) and
\(\mathbf R_i\succ\zero\) set the relative importance of these two
components. The expectation accounts for disturbances in the device
dynamics and random packet deliveries, and \(J_i\) averages the combined
cost over control cycles.

\subsection{Control Cost under Round-Trip Updates}\label{subsec:control-performance}
The control cost in~\eqref{eq:control-cost-metric} depends on the
state and input second moments, whose evolution is governed by the
device dynamics, actuation delay, and round-trip delivery outcomes.
To this end, we first derive the state and input evolution under
command updates and holding, and then establish the stability
condition and calculate the long-term cost \(J_i(p_i)\), where
\(p_i\) denotes the timely round-trip update probability.

\subsubsection{State and Input Evolution under Wireless Updates}
Under the frame protocol, the round-trip delivery outcome determines
whether the actuator updates or holds its input. Specifically, at
\(t_k=kT\), device \(i\) uploads its sampled state \(\mathbf x_{i,k}\)
to the edge controller, which computes the feedback command
\(\mathbf K_i\mathbf x_{i,k}\).
At \(t_k+d\), where \(d=\tau^{\rm u}+\tau^{\rm e}+\tau^{\rm d}\leq T\),
the actuator applies this command if both the UL state and DL command
packets arrive on time, and otherwise retains its previous input.
The applied input in~\eqref{eq:control-cost-metric} therefore follows
the hold-input rule~\cite{schenato2009hold}
\begin{equation}\label{eq:hold-command}
 \mathbf u_{i,k}=\delta_{i,k}\mathbf K_i\mathbf x_{i,k}
 +(1-\delta_{i,k})\mathbf u_{i,k-1}.
\end{equation}
Here, \(\delta_{i,k}\) equals one when both packets arrive on time and
zero otherwise, with \(p_i=\Pr\{\delta_{i,k}=1\}\). The prescribed
feedback gain \(\mathbf K_i\) stabilizes the delayed system when every
round trip succeeds.
To evaluate the state contribution to~\eqref{eq:control-cost-metric},
we propagate the state with \(\mathbf u_{i,k-1}\) over the first \(d\)
seconds and \(\mathbf u_{i,k}\) over the remaining \(T-d\) seconds of
each cycle. Accounting for this actuation delay gives the
sampled dynamics~\cite{nilsson1998stochastic}
\begin{equation}\label{eq:plant-dynamics}
 \mathbf x_{i,k+1}=\mathbf A_i\mathbf x_{i,k}
 +\mathbf B_{i,0}\mathbf u_{i,k}
 +\mathbf B_{i,1}\mathbf u_{i,k-1}+\mathbf w_{i,k}.
\end{equation}
Here, \(\mathbf w_{i,k}\) is the disturbance accumulated over the control
cycle. For the continuous-time state and input matrices
\((\mathbf A_i^{\rm c},\mathbf B_i^{\rm c})\), the sampled coefficients are
\(\mathbf A_i=e^{\mathbf A_i^{\rm c}T}\),
\(\mathbf B_{i,0}=\int_d^T e^{\mathbf A_i^{\rm c}(T-t)}
\mathbf B_i^{\rm c}\,\mathrm dt\), and
\(\mathbf B_{i,1}=\int_0^d e^{\mathbf A_i^{\rm c}(T-t)}
\mathbf B_i^{\rm c}\,\mathrm dt\).
Since \eqref{eq:hold-command} and \eqref{eq:plant-dynamics} both depend
on \(\mathbf u_{i,k-1}\), evaluating the state and input moments requires
tracking the sampled state and previous input jointly.
With the augmented state
\(\mathbf z_{i,k}=[\mathbf x_{i,k}^{\T},\mathbf u_{i,k-1}^{\T}]^{\T}\),
combining these equations yields
\begin{equation}\label{eq:control-mode-dynamics}
 \mathbf z_{i,k+1}
 =\mathbf F_{i,\delta_{i,k}}\mathbf z_{i,k}
 +\overline{\mathbf w}_{i,k},
\end{equation}
where \(\overline{\mathbf w}_{i,k}
=[\mathbf w_{i,k}^{\T},\zero^{\T}]^{\T}\), and the mode matrix is
\(\mathbf F_{i,\delta}=\bigl[
 \begin{smallmatrix}
 \mathbf A_i+\delta\mathbf B_{i,0}\mathbf K_i&
 \mathbf B_{i,1}+(1-\delta)\mathbf B_{i,0}\\
 \delta\mathbf K_i&(1-\delta)\I
 \end{smallmatrix}\bigr]\).

\subsubsection{Evaluation of the Long-Term Control Cost}
Based on the augmented dynamics in~\eqref{eq:control-mode-dynamics}, we
derive the state and input second moments under delayed actuation and
command holding to obtain the long-term control cost \(J_i(p_i)\).
To evaluate the expectation in~\eqref{eq:control-cost-metric}, we express
both quadratic terms using \(\mathbf z_{i,k}\). Its first block is
\(\mathbf x_{i,k}\), and the applied input is
\(\mathbf u_{i,k}=\mathbf U_{i,\delta_{i,k}}\mathbf z_{i,k}\), where
\(\mathbf U_{i,1}=[\mathbf K_i\ \zero]\) and
\(\mathbf U_{i,0}=[\zero\ \I]\) select the fresh and held inputs,
respectively. Averaging over these two outcomes gives
\begin{equation}\label{eq:expected-stage-cost}
 \E[\mathbf x_{i,k}^{\T}\mathbf Q_i\mathbf x_{i,k}
 +\mathbf u_{i,k}^{\T}\mathbf R_i\mathbf u_{i,k}]=\tr(\overline{\mathbf Q}_i(p)\boldsymbol\Sigma_{i,k}),
\end{equation}
where \(p=p_i\) and
\(\boldsymbol\Sigma_{i,k}=\E[\mathbf z_{i,k}\mathbf z_{i,k}^{\T}]\).
This averaging uses the independence of \(\delta_{i,k}\) and
\(\mathbf z_{i,k}\) for fixed channels and state-independent resources.
Furthermore, the average cost weight \(\overline{\mathbf Q}_i(p)\) combines the
state weight with the input weights of the two delivery outcomes, expressed as
\begin{equation}\label{eq:augmented-cost-matrix}
\begin{aligned}
 \overline{\mathbf Q}_i(p)
 =\operatorname{diag}(\mathbf Q_i,\zero)
 +p\mathbf U_{i,1}^{\T}\mathbf R_i\mathbf U_{i,1}+(1-p)\mathbf U_{i,0}^{\T}\mathbf R_i\mathbf U_{i,0}.
\end{aligned}
\end{equation}
To evaluate \(J_i(p)\), we track \(\boldsymbol\Sigma_{i,k}\) across
control cycles. Each device's updates are i.i.d. Bernoulli with success
probability \(p\) and independent of its preceding history. The process
disturbance has zero mean and covariance \(\mathbf W_i\) conditional on
that history and the current update outcome. Under these assumptions,
taking second moments of~\eqref{eq:control-mode-dynamics} gives
\begin{align}
 \boldsymbol\Sigma_{i,k+1}
 =&\,p\mathbf F_{i,1}\boldsymbol\Sigma_{i,k}\mathbf F_{i,1}^{\T}
 +(1-p)\mathbf F_{i,0}\boldsymbol\Sigma_{i,k}\mathbf F_{i,0}^{\T}
 +\overline{\mathbf W}_i,
 \label{eq:covariance-recursion}
\end{align}
where \(\overline{\mathbf W}_i
=\operatorname{diag}(\mathbf W_i,\zero)\) is the disturbance contribution.
Substituting~\eqref{eq:expected-stage-cost}
into~\eqref{eq:control-cost-metric} gives
\begin{equation}\label{eq:control-cost-moment-average}
 J_i(p)=\limsup_{N_{\rm c}\to\infty}
 \tr\!\left(\overline{\mathbf Q}_i(p)
 \frac{1}{N_{\rm c}}\sum_{k=0}^{N_{\rm c}-1}
 \boldsymbol\Sigma_{i,k}\right).
\end{equation}
On the mean-square stable domain specified in
Remark~\ref{rem:control-cost-convergence},
\(\boldsymbol\Sigma_{i,k}\) converges to the stationary second moment
\(\boldsymbol\Sigma_i(p)\), and its time average has the same limit.
Taking this limit in~\eqref{eq:control-cost-moment-average} gives
\begin{equation}\label{eq:general-control-cost}
 J_i(p)=\tr\!\left(\overline{\mathbf Q}_i(p)
 \boldsymbol\Sigma_i(p)\right).
\end{equation}
The second moment \(\boldsymbol\Sigma_i(p)\)
in~\eqref{eq:general-control-cost} is obtained by setting
\(\boldsymbol\Sigma_{i,k+1}=\boldsymbol\Sigma_{i,k}
=\boldsymbol\Sigma_i(p)\) in~\eqref{eq:covariance-recursion} and solving
the resulting linear matrix equation, which gives
\begin{equation}\label{eq:general-closed-form-covariance}
 \vecop(\boldsymbol\Sigma_i(p))
 =[\I-\mathbf M_i(p)]^{-1}\vecop(\overline{\mathbf W}_i),
\end{equation}
where \(\vecop(\cdot)\) stacks matrix columns and
\begin{equation}\label{eq:second-moment-operator}
 \mathbf M_i(p)=p(\mathbf F_{i,1}\otimes\mathbf F_{i,1})
 +(1-p)(\mathbf F_{i,0}\otimes\mathbf F_{i,0}),
\end{equation}
with \(\otimes\) denoting the Kronecker product.

\begin{remark}[Mean-square stability]\label{rem:control-cost-convergence}
The augmented system in~\eqref{eq:control-mode-dynamics} is mean-square
stable if and only if~\cite{costa1993stability,schenato2009hold}
\begin{equation}\label{eq:mean-square-stability}
 \rho(\mathbf M_i(p))<1,
\end{equation}
where \(\rho(\cdot)\) is the spectral radius. Here, mean-square stability
means \(\E[\|\mathbf z_{i,k}\|^2]\to0\) in the absence of process
disturbances, for every finite initial second moment. When every round
trip succeeds (\(p=1\)), the prescribed feedback gain \(\mathbf K_i\)
stabilizes the delayed system, giving
\(\rho(\mathbf M_i(1))=\rho(\mathbf F_{i,1})^2<1\).
For any update probability \(p\) that violates
\eqref{eq:mean-square-stability}, we define \(J_i(p)=+\infty\).
\end{remark}
\begin{proof}
Please refer to Appendix~\ref{app:control-cost-proof}.
\end{proof}

\subsection{Communication--Control Performance Region}
\label{subsec:performance-region}
\label{subsec:round-trip-reliability}
This subsection constructs an achievable communication--control
performance region and characterizes its frame-average throughput
boundary under prescribed control-cost limits. Specifically, we first define the
original performance region, then reformulate its cost constraints as
sufficient reliability and SINR requirements to construct an achievable
inner region.

The communication--control performance region is
\begin{equation}\label{eq:original-performance-region}
\begin{aligned}
 \mathscr R=\bigcup\nolimits_{\boldsymbol\varphi\in\mathcal X_{\rm C}^{\rm hw},\,
 \boldsymbol\Theta\in\mathcal T}
 \bigl\{(\overline{\mathbf J},C):\;&J_i(p_i)\leq\overline J_i,\ \forall i,\\[-1pt]
 &0\leq C\leq C(\boldsymbol\varphi,\boldsymbol\Theta)\bigr\},
\end{aligned}
\end{equation}
where \(\overline{\mathbf J}=[\overline J_1,\ldots,\overline J_D]^{\T}\)
collects the device cost limits and
\(p_i=p_i(\boldsymbol\varphi,\boldsymbol\Theta)\) is the round-trip update
probability. A point \((\overline{\mathbf J},C)\) is feasible when one
admissible design supports frame-average throughput of at least
\(C\) while keeping every device's control cost within its prescribed
limit. We next analyze the control-cost and data-throughput constraints and combine the resulting conditions into an achievable inner
region.

\begingroup
\setlength{\parindent}{0pt}
\noindent\(\bullet\) \textit{Control-cost constraint.}
The cost constraints in~\eqref{eq:original-performance-region} involve
stationary control moments, making direct optimization over communication
resources difficult. To solve this, we first derive a sufficient round-trip reliability
target for each device, then translate its UL/DL allocation into
directional SINR constraints to construct an achievable inner region.

For device \(i\), let \(p_i^{\rm req}(\overline J_i)\) be the supremum
of update probabilities that violate stability or
\(J_i(p)\leq\overline J_i\), taken as zero if none violate. When
\(p_i^{\rm req}<1\), choose
\(p_i^{\rm req}<\widehat p_i(\overline J_i)<1\). Every
\(p_i\geq\widehat p_i\) then satisfies both requirements, reducing the
control-cost constraint to a sufficient round-trip reliability requirement.

To impose this requirement on the UL and DL SINRs, we use their decoding
success probabilities. Under the finite-blocklength normal
approximation~\cite{polyanskiy2010finite}, the directional success probability is
\(\psi_i^a(\gamma)=1-\epsilon_{i,\rm NA}^a(\gamma)\), with
\begin{equation}\label{eq:fbl-error}
 \epsilon_{i,\rm NA}^a
 \triangleq\mathrm Q\!\left(
 \frac{n_i^a\log_2(1+\gamma_i^{\mathrm{ctrl},a})-L_i^a}
 {\sqrt{n_i^aV(\gamma_i^{\mathrm{ctrl},a})}}
 \right),
\end{equation}
where \(L_i^a\) is the packet length, \(n_i^a=B^a\tau^a\) the blocklength,
\(\mathrm Q(\cdot)\) the Gaussian tail function, and
\(V(\gamma)=[1-(1+\gamma)^{-2}](\log_2e)^2\) the dispersion for
additive white Gaussian noise. For fixed channels and state-independent resources, independent
receiver noise and interfering symbols in the separate UL and DL
intervals give the round-trip update probability
\begin{equation}\label{eq:timely-probability}
 p_i(\boldsymbol\varphi,\boldsymbol\Theta)
 =\psi_i^{\rm u}(\gamma_i^{\mathrm{ctrl,u}})
  \psi_i^{\rm d}(\gamma_i^{\mathrm{ctrl,d}}).
\end{equation}
Here, the DL success probability is conditional on a successful upload.
By~\eqref{eq:timely-probability}, the required round-trip reliability is
guaranteed if the two links meet minimum success targets
\(s_i^{\rm u}\) and \(s_i^{\rm d}\) satisfying
\[
 s_i^{\rm u}s_i^{\rm d}\geq\widehat p_i,\qquad
 \psi_i^a(\gamma_i^{\mathrm{ctrl},a})\geq s_i^a,\quad\forall a.
\]
For positive packet lengths, \(\psi_i^a\) increases strictly with SINR.
Each link-success requirement can therefore be imposed as
\(\gamma_i^{\mathrm{ctrl},a}\geq\Gamma_i^a(s_i^a)\), where
\(\Gamma_i^a\) is defined in~\eqref{eq:inverse-fbl}.
Let \(\mathcal S_i(\overline J_i)\) contain target pairs
\(\mathbf s_i=(s_i^{\rm u},s_i^{\rm d})\) satisfying the product
requirement and prescribed ranges
\(\underline s_i^a\leq s_i^a\leq\overline s_i^a\), with
\(0<\underline s_i^a\leq\overline s_i^a<1\).

\noindent\(\bullet\) \textit{Data-throughput constraint.}
For the same admissible joint design satisfying these SINR constraints,
the frame-average data throughput is given
by~\eqref{eq:communication-utility}. A throughput level \(C\) is therefore
supported whenever
\(0\leq C\leq C(\boldsymbol\varphi,\boldsymbol\Theta)\).
\endgroup

Combining the sufficient control-cost constraints with this throughput
condition over all admissible joint designs yields the achievable inner
region
\(\mathscr R_{\rm in}\) in~\eqref{eq:inner-performance-region}.
Each admitted design satisfies
\(p_i\geq s_i^{\rm u}s_i^{\rm d}\geq\widehat p_i\) and hence
\(J_i(p_i)\leq\overline J_i\), proving
\(\mathscr R_{\rm in}\subseteq\mathscr R\). The inclusion can be
strict because the sufficient reliability targets and their prescribed
directional ranges may exclude designs satisfying the original cost
constraints.
The inner region thus provides sufficient communication constraints
for joint resource design under prescribed control-cost limits in the subsequent section.

\section{Joint C-PASS and Beamforming Optimization}\label{sec:design}
Building on the preceding performance characterization, this section
develops a joint C-PASS and transceiver design for data throughput
maximization under prescribed control-cost limits. Specifically, we
formulate the joint design problem, reformulate its control-cost
constraints into sufficient directional SINR requirements, and
decompose the resulting resource optimization. This decomposition
leads to an alternating optimization (AO) algorithm that updates
reliability and transceiver resources and reconfigures C-PASS with
resource reoptimization. Finally, we summarize the algorithm and analyze
its convergence and computational complexity.

\subsection{Problem Formulation}
The performance region in~\eqref{eq:original-performance-region}
characterizes feasible combinations of control-cost limits and data
throughput, but leaves the corresponding resource design unspecified.
The C-PASS configuration determines both control and data channels,
coupling control protection and data access. This subsection formulates
a joint C-PASS and transceiver design problem to maximize data
throughput under prescribed control-cost limits.
Specifically, the joint design problem \(\mathcal P_0\) is formulated as
\begin{subequations}\label{prob:original-joint-design}
\begin{align}
 C_{\rm orig}^{\star}(\overline{\mathbf J})
 =\sup_{\boldsymbol\varphi,\boldsymbol\Theta}\quad&
 C(\boldsymbol\varphi,\boldsymbol\Theta)\\
 \mathrm{s.t.}\quad&
 J_i\!\left(p_i(\boldsymbol\varphi,\boldsymbol\Theta)\right)
 \leq\overline J_i,\quad i\in\mathcal D,\\
 &\boldsymbol\Theta\in\mathcal T,\quad
 \boldsymbol\varphi\in\mathcal X_{\rm C}^{\rm hw},
\end{align}
\end{subequations}
where \(\overline{\mathbf J}=[\overline J_1,\ldots,\overline J_D]^{\T}\)
collects the prescribed control-cost limits.
In~\eqref{prob:original-joint-design},
\(C_{\rm orig}^{\star}(\overline{\mathbf J})\) denotes the supremum
of the frame-average weighted data throughput over the C-PASS
configuration \(\boldsymbol\varphi\) and digital resources
\(\boldsymbol\Theta\). The first constraint requires each device's
control cost, induced by the resulting round-trip update probability,
to remain within its prescribed limit. The remaining constraints
enforce the power and combiner requirements in \(\mathcal T\) and
physical hardware limits in \(\mathcal X_{\rm C}^{\rm hw}\).

However, \(\mathcal P_0\) is difficult to solve directly owing to
its nonconvexity and strong variable coupling, which arise from
1) the nonlinear control-cost expression
\(J_i(p_i(\boldsymbol\varphi,\boldsymbol\Theta))\), which involves
stationary control moments and requires mean-square stability for finiteness;
2) interference coupling among powers, combiners, and beamformers; and
3) coupling between the C-PASS configuration and transceiver resources
in both desired-signal and interference powers.

\subsection{Problem Reformulation and Decomposition}
\label{subsec:design-reformulation}
To address these challenges, this subsection first reformulates the
control-cost constraints and then decomposes the resulting resource
optimization. Specifically, the constraint reformulation and subsequent
resource decomposition are presented as follows.

\subsubsection{Reformulation of Control-Cost Constraints}\label{app:fbl-monotonicity}
To handle the nonlinear control-cost expression, we replace the cost
constraints in~\eqref{prob:original-joint-design} with the sufficient
conditions in~\eqref{eq:inner-performance-region},
\(\mathbf s_i\in\mathcal S_i(\overline J_i)\) and
\(\gamma_i^{\mathrm{ctrl},a}\geq\Gamma_i^a(s_i^a)\) for all
\(i\in\mathcal D\), \(a\in\mathcal A\).
The SINR thresholds \(\Gamma_i^a\) are defined by
\begin{equation}\label{eq:inverse-fbl}
 \Gamma_i^a(s)=\inf\{\gamma\geq0:\psi_i^a(\gamma)\geq s\}.
\end{equation}
For the decoding probability defined by~\eqref{eq:fbl-error}, write
\(\psi(\gamma)=\Phi(t(x))\), where \(\Phi\) is the standard normal
CDF, \(x=1+\gamma>1\), and
\(t(x)=\sqrt n(\ln x-b)/\sqrt{1-x^{-2}}\) with \(b=L\ln2/n>0\).
Differentiation gives \(t'(x)>0\) because
\(x^2-1-\ln x+b>0\) for \(x>1\). Thus, \(\psi\) increases strictly
from zero to one and has a unique inverse on \((0,1)\).
We set \(\psi_i^a(0)=0\) by continuous extension.
For prescribed cost limits, \(\widehat p_i(\overline J_i)\) is determined
from the control analysis. The requirements on \(\mathbf s_i\) and the
directional SINRs guarantee \(p_i\geq\widehat p_i\), so stationary control
moments need not be evaluated during resource optimization. The throughput
objective and transceiver and hardware feasibility sets
in~\eqref{prob:original-joint-design} are unchanged.
Then, the corresponding inner region is
\begin{equation}\label{eq:inner-performance-region}
\begin{aligned}
 \mathscr R_{\rm in}={}&
 \bigcup\nolimits_{\boldsymbol\varphi\in\mathcal X_{\rm C}^{\rm hw},\,
                    \boldsymbol\Theta\in\mathcal T}
 \bigl\{(\overline{\mathbf J},C):\,
 \exists\,\mathbf s_i\in\mathcal S_i(\overline J_i),\ \forall i,\\[-1pt]
 &\gamma_i^{\mathrm{ctrl},a}\geq\Gamma_i^a(s_i^a),\ \forall i,a,\quad
 0\leq C\leq C(\boldsymbol\varphi,\boldsymbol\Theta)\bigr\}.
\end{aligned}
\end{equation}

\subsubsection{Decomposition of the Resource Optimization Problem}
\label{subsubsec:variable-reduction}\label{subsubsec:ul-combining}
To address the remaining variable coupling, we reduce the reliability
variables and eliminate UL combiners analytically, then decompose resource
optimization with a fixed C-PASS configuration and fixed UL/DL reliability targets.

\begingroup
\setlength{\parindent}{0pt}
\noindent\(\bullet\) \textit{Reliability variable reduction and UL combiner elimination.}
To simplify the coupled UL/DL reliability design, we reduce the two targets
to one allocation variable per device. The target product can be reduced until the round-trip
requirement or both lower bounds are active, giving
\begin{equation}\label{eq:effective-reliability-product}
 s_i^{\rm u}s_i^{\rm d}=\widetilde p_i,
\end{equation}
where \(\widetilde p_i=\max\{\widehat p_i,
\underline s_i^{\rm u}\underline s_i^{\rm d}\}\).
If this exceeds \(\overline s_i^{\rm u}\overline s_i^{\rm d}\), the
selected coding range is infeasible. Otherwise, the parameterization
\(s_i^{\rm u}=e^{z_i}\),
\(s_i^{\rm d}=\widetilde p_i e^{-z_i}\) removes the product constraint
and leaves one bounded variable \(z_i\in\mathcal Z_i\) per device.
The directional coding bounds give
\begin{equation}\label{eq:reliability-search-interval}
\begin{aligned}
 \mathcal Z_i=\bigl[
 \ln\max\{\underline s_i^{\rm u},
 \widetilde p_i/\overline s_i^{\rm d}\},\ln\min\{\overline s_i^{\rm u},
 \widetilde p_i/\underline s_i^{\rm d}\}\bigr].
\end{aligned}
\end{equation}

After reducing the reliability variables, we further simplify resource
optimization by eliminating UL combiners analytically. At fixed powers
and C-PASS configuration, each combiner is optimized independently to
maximize its own SINR. The resulting minimum mean-square error (MMSE) direction supports throughput
and control feasibility, expressed as
\begin{equation}\label{eq:mmse-control-combiner}
 \mathbf v_r=\frac{\mathbf R_{-r}^{-1}\mathbf h_r^{\rm u}}
 {\|\mathbf R_{-r}^{-1}\mathbf h_r^{\rm u}\|},
\end{equation}
where \(\mathbf R_{-r}=\sigma_{\rm u}^2\I_M+
\sum_{\ell\ne r}P_\ell\mathbf h_\ell^{\rm u}
(\mathbf h_\ell^{\rm u})^{\Hh}\).
Let \(\boldsymbol\Theta_{\rm tx}=(\mathbf P,\mathbf W,\mathbf P_0,\mathbf W_0)\) collect
concurrent and data-only powers and DL beams. Substituting the reliability
parameterization and MMSE combiners into the sufficient SINR reformulation
of \(\mathcal P_0\) gives the following problem \(\mathcal P_1\)
over \(\boldsymbol\varphi\), \(\boldsymbol\Theta_{\rm tx}\), and \(\mathbf z\),
\begin{subequations}\label{prob:reduced-joint-design}
\begin{align}
 C_{\rm C}^{\star}(\overline{\mathbf J})
 &=\max_{\boldsymbol\varphi,\boldsymbol\Theta_{\rm tx},\mathbf z}\;
 C(\boldsymbol\varphi,\boldsymbol\Theta_{\rm tx})\\
 \mathrm{s.t.}\quad&
 \gamma_i^{\mathrm{ctrl},a}\geq\Gamma_i^a(s_i^a(z_i)),
 \quad i\in\mathcal D,\ a\in\mathcal A,
 \label{eq:reduced-control-uplink}\\
 &\boldsymbol\Theta_{\rm tx}\in\mathcal T_{\rm tx},\quad z_i\in\mathcal Z_i,
 \quad i\in\mathcal D,\label{eq:reduced-transmit-resources}\\
 &\boldsymbol\varphi\in\mathcal X_{\rm C}^{\rm hw}.
\end{align}
\end{subequations}
Here, \(\mathcal T_{\rm tx}\) is obtained from \(\mathcal T\) by
eliminating the UL combiner variables. It comprises power and beamforming
allocations satisfying the original UL power bounds and DL sum-power
limits for concurrent and data-only transmission.

\noindent\(\bullet\) \textit{Decomposition of resource optimization.}
For a fixed C-PASS configuration \(\overline{\boldsymbol\varphi}\),
the resource subproblem of~\eqref{prob:reduced-joint-design} separates
into concurrent and data-only UL/DL subproblems. With the UL/DL
reliability targets also fixed, their throughput contributions are
additive and their resource constraints are independent across
transmission intervals. Let \(C_0^\star\) be the optimal total data-only
contribution and \(C_{\rm co}^{a,\star}\) the optimal concurrent
contribution under \(\mathbf s^a=[s_i^a]_{i\in\mathcal D}\), including
the durations, bandwidths, and user weights
in~\eqref{eq:communication-utility}. When the sufficient constraints are feasible, the
fixed-hardware throughput optimum decomposes as
\begin{equation}\label{eq:inner-boundary-decomposition}
 C_{\rm tx}^{\rm ach}(\overline{\mathbf J};\overline{\boldsymbol\varphi})
 =C_0^\star+
 \max_{\{z_i\in\mathcal Z_i\}_{i\in\mathcal D}}
 \sum_{a\in\mathcal A}C_{\rm co}^{a,\star}(\mathbf s^a(\mathbf z)),
\end{equation}
where \(\mathbf s^a(\mathbf z)=[s_i^a(z_i)]_{i\in\mathcal D}\) and the fixed
\(\overline{\boldsymbol\varphi}\) argument is omitted on the right-hand side.
The concurrent UL/DL resources
remain coupled through \(\mathbf z\), while C-PASS reconfiguration affects
the channels of all resource subproblems. These remaining couplings
motivate the AO procedure developed in the next subsection.

\par
\endgroup

\subsection{AO-Based Joint Optimization}\label{subsec:ao-design}
\label{app:resource-updates}
Building on the decomposition
in~\eqref{eq:inner-boundary-decomposition}, we develop an AO
procedure for the reduced joint design problem
\(\mathcal P_1\) in~\eqref{prob:reduced-joint-design}.
Specifically, we optimize reliability and transceiver resources
at a fixed C-PASS configuration. The resulting allocation and
control multipliers guide the C-PASS configuration update,
with each candidate evaluated after resource reoptimization.

\subsubsection{Joint Reliability and Transceiver Optimization}
\label{subsubsec:coexistence-update}
\label{subsubsec:data-only-update}
\label{subsec:transceiver-updates}
For a given C-PASS configuration \(\boldsymbol\varphi^{(k)}\), we jointly
optimize reliability allocation and concurrent powers and beams because
they are coupled through the control-SINR constraints. We then solve the
data-only UL and DL subproblems independently, exploiting their additive
throughput contributions and separate power budgets. Together, these
updates provide the resource allocation and control multipliers used in
the subsequent C-PASS configuration update.

\begingroup
\setlength{\parindent}{0pt}
\noindent\(\bullet\) \textit{Joint concurrent resource and reliability update.}
With \(\boldsymbol\varphi=\boldsymbol\varphi^{(k)}\) and feasible data-only
resources \((\mathbf P_0,\mathbf W_0)\) held fixed, the concurrent block
of~\eqref{prob:reduced-joint-design} is
\begin{equation}\label{prob:concurrent-resource-update}
\begin{aligned}
 \max_{\mathbf P,\mathbf W,\mathbf z}\quad&
 C(\boldsymbol\varphi^{(k)},\boldsymbol\Theta_{\rm tx})\\
 \mathrm{s.t.}\quad&
 \gamma_i^{\mathrm{ctrl},a}\geq\Gamma_i^a(s_i^a(z_i)),
 \quad i\in\mathcal D,\ a\in\mathcal A,\\
 &0\leq P_r\leq P_{r,\max}^{\rm u},\quad r\in\mathcal R,\\
 &\|\mathbf W\|_{\rm F}^2\leq P_{\max}^{\rm d},\quad
 z_i\in\mathcal Z_i,\ i\in\mathcal D.
\end{aligned}
\end{equation}
Here, \(\boldsymbol\Theta_{\rm tx}=(\mathbf P,\mathbf W,\mathbf P_0,\mathbf W_0)\),
so the data-only throughput terms are constant in this block. The C-PASS
constraint is already satisfied by \(\boldsymbol\varphi^{(k)}\), and the
remaining constraints are inherited from
\eqref{eq:reduced-control-uplink}--\eqref{eq:reduced-transmit-resources}.
For numerical solution, we define the normalized negative throughput
\(F=-C/B_{\rm ref}\), where \(B_{\rm ref}>0\) is a fixed normalization
constant. Minimizing \(F\) is equivalent to maximizing \(C\) over the
same feasible set. With the C-PASS configuration and data-only resources
fixed, we write \(F(\boldsymbol\eta)\), where \(\boldsymbol\eta\) collects
the scaled real coordinates of \((\mathbf P,\mathbf W,\mathbf z)\).
In these coordinates, the constraints are expressed as normalized
residuals \(\mathbf c(\boldsymbol\eta)\leq\zero\), with control entries
\(c_i^a=1-\gamma_i^{\mathrm{ctrl},a}/\Gamma_i^a(s_i^a)\).

The objective and constraints in~\eqref{prob:concurrent-resource-update}
are smooth but nonconvex because of interference coupling and
reliability-dependent SINR thresholds. We therefore apply sequential
quadratic programming (SQP)~\cite{nocedal2006numerical} in scaled real
coordinates. This yields joint updates of powers, beams, and reliability,
together with the control multipliers \(\lambda_i^a\geq0\) required by
the subsequent C-PASS configuration step. Specifically, at inner iteration \(t\), we use
analytical derivatives at \(\boldsymbol\eta_t^{(k)}\) to form a quadratic
Lagrangian model with linearized constraints and obtain the joint
direction \(\mathbf d_t^{(k)}\) by solving
\begin{equation}\label{prob:concurrent-sqp-step}
\begin{aligned}
 \min_{\mathbf d}\quad&
 \nabla F(\boldsymbol\eta_t^{(k)})^{\T}\mathbf d
 +\tfrac12\mathbf d^{\T}\mathbf H_t^{(k)}\mathbf d\\
 \mathrm{s.t.}\quad&
 \mathbf c(\boldsymbol\eta_t^{(k)})+\mathbf J_t^{(k)}\mathbf d\leq\zero,
\end{aligned}
\end{equation}
where \(\mathbf H_t^{(k)}\) approximates the Lagrangian Hessian and
\(\mathbf J_t^{(k)}\) is the constraint Jacobian at
\(\boldsymbol\eta_t^{(k)}\). The resulting direction updates the variables
as \(\boldsymbol\eta_{t+1}^{(k)}=\boldsymbol\eta_t^{(k)}+
\alpha_t^{(k)}\mathbf d_t^{(k)}\), with \(\alpha_t^{(k)}\) selected by
line search. Since this subproblem uses linearized constraints, the
final resource allocation is checked against the original nonlinear
constraints.

The initial DL beams are obtained by a fixed-target weighted MMSE
(WMMSE) step~\cite{shi2011wmmse} with second-order cone control-SINR
constraints and the BS power budget.
The required reliability sensitivities are
\(\dot\Gamma_i^{\rm u}=s_i^{\rm u}/[\psi_i^{\rm u}]'(\Gamma_i^{\rm u})\)
and
\(\dot\Gamma_i^{\rm d}=-s_i^{\rm d}/[\psi_i^{\rm d}]'(\Gamma_i^{\rm d})\),
where \(\dot\Gamma_i^a=\mathrm d\Gamma_i^a/\mathrm dz_i\) and
\(\Gamma_i^a=\Gamma_i^a(s_i^a)\).

\noindent\(\bullet\) \textit{Data-only UL power update.}
With \(\boldsymbol\varphi^{(k)}\) fixed, data-only UL power optimization
is independent of the other resource blocks. After MMSE combiner
elimination, the power subproblem is
\begin{equation}\label{prob:data-only-uplink}
 \max_{\{0\leq P_{D+j}^0\leq P_{D+j,\max}^{\rm u}\}_{j\in\mathcal K}}
 C_0^{\rm u},
\end{equation}
where \(C_0^{\rm u}\) is the data-only UL contribution
in~\eqref{eq:communication-utility}.
Interference makes the objective generally nonconcave, so we apply the
preceding SQP procedure to the UL power coordinates. Specifically,
analytical derivatives of the MMSE-substituted throughput define a
quadratic power model subject to the bounds. A line search then
determines the update, with MMSE combiners recomputed at each trial.
Only feasible candidates that increase \(C_0^{\rm u}\) replace the incumbent.

\noindent\(\bullet\) \textit{Data-only DL beam update.}
With the C-PASS configuration fixed, the data-only DL beamforming
subproblem is
\begin{equation}\label{prob:data-only-downlink}
 \max_{\|\mathbf W_0\|_{\rm F}^2\leq P_{\max}^{\rm d}}
 C_0^{\rm d},
\end{equation}
where \(C_0^{\rm d}\) is the data-only DL contribution
in~\eqref{eq:communication-utility}.
Although interference makes this objective generally nonconcave,
WMMSE yields a convex quadratic beam subproblem at fixed auxiliary
receivers and weights. Its single sum-power constraint allows the beam
update to be computed through a scalar power multiplier.

Specifically, for the current beams, define
\(a_{0,j\ell}=(\mathbf h_{D+j}^{\rm d})^{\Hh}\mathbf w_{D+\ell}^0\) and
\(t_{0,j}=\sum_{\ell\in\mathcal K}|a_{0,j\ell}|^2+\sigma_{D+j}^2\).
The auxiliary updates are \(u_{0,j}=a_{0,jj}/t_{0,j}\) and
\(\beta_{0,j}=e_{0,j}^{-1}=1+\gamma_j^{\mathrm{com,d},0}\), where
\(e_{0,j}\) is the resulting minimum MSE.
At fixed receivers and weights, the minimum-norm beam update is
\begin{equation}\label{eq:data-only-wmmse-scalar-dual}
 \mathbf W_0(\mu)=(\mathbf A_0+\mu\I)^{\dagger}\mathbf G_0,
 \quad \mu\geq0,
\end{equation}
where
\(\mathbf A_0=\sum_{j\in\mathcal K}\nu_{0,j}\beta_{0,j}|u_{0,j}|^2
\mathbf h_{D+j}^{\rm d}(\mathbf h_{D+j}^{\rm d})^{\Hh}\),
\(\mathbf G_0=[\mathbf b_{0,r}]_{r\in\mathcal R_{\rm d}}\), and
\(\mathbf b_{0,D+j}=\nu_{0,j}\beta_{0,j}u_{0,j}\mathbf h_{D+j}^{\rm d}\),
with \(\nu_{0,j}=\omega_jB^{\rm d}\tau_0^{\rm d}/(T\ln2)\).
Here, \(\dagger\) denotes the Moore--Penrose pseudoinverse and
\(\mu\) is the multiplier associated with the BS power budget.
The receiver, weight, and beam updates are repeated until the relative
throughput gain meets the prescribed tolerance or a computation limit
is reached.

\par
\endgroup

\subsubsection{C-PASS Configuration Update with Transceiver Reoptimization}
\label{subsubsec:hardware-update}
C-PASS reconfiguration changes desired gains and interference, so its
throughput benefit depends on the accompanying resource adaptation.
We therefore use a local Lagrangian model to generate configuration
directions and evaluate them after resource reoptimization.
Specifically, at \(\boldsymbol\varphi^{(k)}\), the current resource
solution and control multipliers give the scaled sensitivity
\begin{equation}\label{eq:physical-value-gradient}
 \mathbf g_k=\mathbf S_{\varphi}
 \left(\nabla_{\boldsymbol\varphi}F+
 \sum_{i,a}\lambda_i^a\nabla_{\boldsymbol\varphi}c_i^a\right),
\end{equation}
where the positive diagonal matrix \(\mathbf S_{\varphi}\) defines the
scaled coordinates \(\boldsymbol\chi=\mathbf S_{\varphi}^{-1}\boldsymbol\varphi\).
The first term captures the sensitivity of \(F\) with the resource
allocation held fixed, while the multiplier terms account for changes
in the control-SINR margins.

To include curvature without full second derivatives, we update
\(\mathbf B_k\succ0\) from successive sensitivities using Powell-damped
BFGS~\cite{nocedal2006numerical} with positive diagonal loading.
The direction \(\mathbf d_k\) then solves
\begin{equation}\label{prob:cpass-sca-update}
\begin{aligned}
 \min_{\mathbf d}\quad&\mathbf g_k^{\T}\mathbf d+
 \tfrac12\mathbf d^{\T}\mathbf B_k\mathbf d\\
 \mathrm{s.t.}\quad&\boldsymbol\varphi^{(k)}+
 \mathbf S_{\varphi}\mathbf d\in\mathcal X_{\rm C}^{\rm hw},
 \qquad|\mathbf d|\leq\mathbf r_k.
\end{aligned}
\end{equation}
The retained C-PASS constraints, including passivity, remain nonlinear,
so we solve this model by SQP using analytical derivatives, with
\(\mathbf r_k\) bounding coordinate changes.

For each candidate
\(\boldsymbol\varphi^{\rm tr}=\boldsymbol\varphi^{(k)}+
\alpha\mathbf S_{\varphi}\mathbf d_k\), we recompute the channels and
reoptimize reliability and transceiver resources. We update the current
design only when the candidate satisfies the original constraints and
improves throughput. Otherwise, backtracking reduces \(\alpha\) and
repeats this evaluation.

\subsection{Overall Algorithm and Analysis}
Algorithm~\ref{alg:boundary-computation} summarizes the proposed AO
procedure, which alternates between reliability and transceiver
optimization and C-PASS configuration updates. We next analyze its
convergence properties and computational complexity.

\begin{algorithm}[!t]
\caption{AO for Joint C-PASS and Transceiver Design}
\label{alg:boundary-computation}
\renewcommand{\algorithmicrequire}{\textbf{Input:}}
\renewcommand{\algorithmicensure}{\textbf{Output:}}
\begin{algorithmic}[1]
\REQUIRE System parameters, cost limits \(\overline{\mathbf J}\), a feasible
\((\boldsymbol\varphi,\boldsymbol\Theta_{\rm tx},\mathbf z)\), and stopping parameters
\ENSURE Feasible \((\boldsymbol\varphi,\mathbf s,\boldsymbol\Theta)\) and throughput \(C_{\rm ALG}\)
\STATE Compute the reliability bounds and \(\mathcal Z_i\), \(\forall i\in\mathcal D\).
Initialize \(k=0\), \(\mathbf B_0\succ0\), and \(\mathbf r_0\).
\REPEAT
\STATE \label{algline:concurrent-update}At fixed \(\boldsymbol\varphi\), update
\((\mathbf P,\mathbf W,\mathbf z)\) by SQP for
\eqref{prob:concurrent-resource-update} and retain \(\lambda_i^a\).
\STATE Update \(\mathbf P_0\) by SQP for \eqref{prob:data-only-uplink}.
\STATE \label{algline:downlink-update}Update \(\mathbf W_0\) by WMMSE for
\eqref{prob:data-only-downlink}.
\STATE Compute \(\mathbf g_k\), update \(\mathbf B_k\), and solve
\eqref{prob:cpass-sca-update} by SQP for \(\mathbf d_k\).\par
\STATE Perform backtracking with
\(\boldsymbol\varphi^{\rm tr}=\boldsymbol\varphi+
\alpha\mathbf S_{\varphi}\mathbf d_k\), reoptimizing
\((\boldsymbol\Theta_{\rm tx}^{\rm tr},\mathbf z^{\rm tr})\) at each trial.
\IF{the trial is feasible for \eqref{prob:reduced-joint-design} and improves \(C\)}
\STATE \((\boldsymbol\varphi,\boldsymbol\Theta_{\rm tx},\mathbf z)\leftarrow
(\boldsymbol\varphi^{\rm tr},\boldsymbol\Theta_{\rm tx}^{\rm tr},\mathbf z^{\rm tr})\).
\ENDIF
\STATE Adjust \(\mathbf r_k\) and set \(k\leftarrow k+1\).
\UNTIL{the prescribed stopping criterion is satisfied or a computational limit is reached}
\STATE Recover \(\mathbf s\) and the UL receivers, and return
\((\boldsymbol\varphi,\mathbf s,\boldsymbol\Theta)\) and
\(C_{\rm ALG}=C(\boldsymbol\varphi,\boldsymbol\Theta)\).
\end{algorithmic}
\end{algorithm}
\subsubsection{Convergence Analysis}\label{subsubsec:ao-convergence}
Starting from a feasible design, the algorithm accepts only feasible
updates that do not decrease throughput. Let \(C^{(k)}\) denote
the throughput at the beginning of iteration \(k\), and let
\(\widetilde C^{(k)}\) denote the throughput after the resource updates.
The resource and configuration updates then give
\begin{equation}\label{eq:ao-throughput-monotonicity}
 C^{(k)}\leq\widetilde C^{(k)}\leq C^{(k+1)},\qquad k\geq0,
\end{equation}
where the second inequality holds with equality if no configuration
trial is accepted. Moreover, bounded channels and transmit powers
with positive receiver noise impose a finite upper bound on throughput.
Hence, the nondecreasing sequence \(\{C^{(k)}\}\) converges to a finite
limit. This argument relies on the acceptance rules and does not
require globally optimal solutions to the subproblems.
The throughput achieved by Algorithm~\ref{alg:boundary-computation}
satisfies
\begin{equation}\label{eq:achievable-boundary-value}
 C_{\rm ALG}\leq C_{\rm C}^{\star}(\overline{\mathbf J})
 \leq C_{\rm orig}^{\star}(\overline{\mathbf J}),
\end{equation}
where the first inequality follows from feasibility for
\eqref{prob:reduced-joint-design}, and the second follows from its
sufficient control constraints. Thus, the throughput sequence generated
by the proposed algorithm converges to a finite value.

\subsubsection{Computational Complexity}\label{subsubsec:ao-complexity}
The concurrent resource, data-only UL, data-only DL, and C-PASS blocks
contain \(n_{\rm co}=U(1+2M)+D\), \(K\), \(2MK\), and
\(n_{\varphi}=2N+2M\) real variables, respectively, where \(U=D+K\)
is the total number of terminals. Dense SQP requires
\(\mathcal O(n^3)\) per KKT factorization for an \(n\)-variable block
with \(\mathcal O(n)\) independent active constraints. Including MMSE
receiver updates, the concurrent and data-only UL calls cost
\(\mathcal O(I_{\rm co}n_{\rm co}^3)\) and
\(\mathcal O(I_{\rm u}(M+K)^3)\), respectively, where
\(I_{\rm co}\) and \(I_{\rm u}\) bound the factorization and
function/Jacobian evaluation counts, including line searches.

Each data-only DL WMMSE iteration costs \(\mathcal O(\mathcal C_{\rm w})\),
where \(\mathcal C_{\rm w}=M^3+KM^2+MK^2+L_{\mu}MK\) and
\(L_{\mu}\) counts power-multiplier evaluations. Reusing the
eigendecomposition reduces each multiplier evaluation to
\(\mathcal O(MK)\). Thus, one resource-optimization call costs
\(\mathcal O(\mathcal C_{\rm res})\), where
\(\mathcal C_{\rm res}=I_{\rm co}n_{\rm co}^3+
I_{\rm u}(M+K)^3+I_{\rm w}\mathcal C_{\rm w}\), and \(I_{\rm w}\)
bounds the WMMSE iterations. Channel reconstruction, analytical
sensitivities, and BFGS updates are covered by the dense upper bound
\(\mathcal O(n_{\rm co}^3+n_{\varphi}^3)\).
Accounting for configuration SQP and backtracking, the total
complexity is bounded by
\begin{equation}\label{eq:overall-ao-complexity}
 \mathcal O\!\left(I_{\rm out}\left[
 (1+B)\mathcal C_{\rm res}+(I_{\varphi}+B)n_{\varphi}^3
 \right]\right),
\end{equation}
where \(I_{\rm out}\) is the number of outer iterations, \(B\) bounds
the backtracking trials per iteration, and \(I_{\varphi}\) bounds the
configuration SQP factorization and evaluation counts. The factor
\(1+B\) accounts for resource optimization at the current and trial
configurations, while \(Bn_{\varphi}^3\) covers the additional
configuration checks.

\section{Numerical Results}\label{sec:numerical}
This section validates the control-cost analysis and evaluates the data
throughput supported by C-PASS under prescribed control-cost limits.
We examine AO convergence and compare its throughput with alternative
solvers under a common CPU-time budget. Comparisons with end-fed PASS,
fixed and separately trained hardware, and separated transmission assess
the throughput benefits of the C-PASS architecture, joint hardware
adaptation, and concurrent data and control transmission.

\begin{figure*}[!t]
 \centering
 \subfloat[Control cost.\label{fig:control-mapping-a}]{%
  \includegraphics[width=0.325\textwidth]{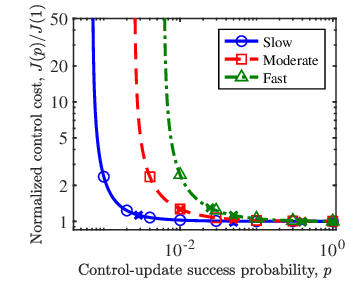}}\hfill
 \subfloat[Required reliability.\label{fig:control-mapping-b}]{%
  \includegraphics[width=0.325\textwidth]{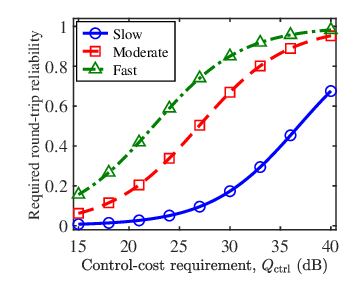}}\hfill
 \subfloat[Required SINR.\label{fig:control-mapping-c}]{%
  \includegraphics[width=0.325\textwidth]{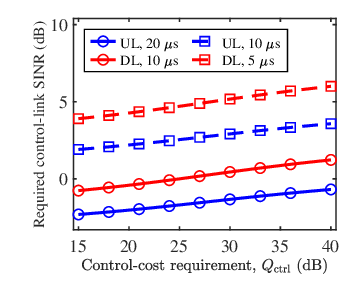}}
 \caption{Control-to-radio mapping. In (a), curves show analytical costs,
 and crosses show Monte Carlo estimates with 95\% confidence intervals.}
 \label{fig:control-mapping}
\end{figure*}

\subsection{Simulation Setup}
\label{sec:sim-setup}
We deploy \(L=2\) waveguides of length \(\ell_l=20\) m at
\((y_1,y_2)=(5,15)\) m and height \(H=3\) m in a square region
with \(L_x=L_y=20\) m. The feed-port and PA coordinates are
\((d_m^{\rm feed},y_l,H)\) and \((q_n,y_l,H)\), respectively.
Each feed pair is fixed at \(d_m^{\rm feed}=\ell_l/2\pm\Delta_{\rm f}/2\),
with center \(\ell_l/2=10\) m and separation
\(\Delta_{\rm f}=5\lambda_{\rm wg}/4\).
Each waveguide carries two PAs on each side of the input region.
Terminal layouts use uniform horizontal candidates over
\([0,L_x]\times[0,L_y]\) at height \(z_r^{\rm t}=1\) m.
Control devices and data users are drawn separately, with minimum
horizontal separations of \(d_{\rm c}^{\min}=2\) m between control
devices and \(d_{\rm t}^{\min}=0.5\) m between all terminal pairs.
A single spherical free-space LoS path determines PA-to-terminal
attenuation and phase from three-dimensional distances.
Compared methods share the same layouts, held fixed during PA adaptation.
Table~\ref{tab:numerical-parameters} lists the common parameters
unless stated otherwise, with PA coordinates and placement intervals
in increasing order on each waveguide.
The frame durations follow the prescribed protocol, with data-only
access including the computation interval. All data-user weights
\(\omega_j\) are \(1\), and individual data streams may be muted.

\begin{table}[!t]
 \caption{Common simulation parameters.}
 \label{tab:numerical-parameters}
 \centering
 \footnotesize
 \setlength{\tabcolsep}{3pt}
 \renewcommand{\arraystretch}{1.08}
 \begin{tabular}{@{}p{0.43\columnwidth}p{0.53\columnwidth}@{}}
  \toprule
  Parameter & Value \\
  \midrule
  RF chains / PAs & \(M=4,\ N=8\) \\
  Control devices / data users & \(D=4,\ K=8\) \\
  Nominal PA coordinates \(q_n^{\rm nom}\) &
  \([2.5,7.5,12.5,17.5]~\mathrm{m}\) \\
  PA intervals \([q_n^{\min},q_n^{\max}]\) &
  \(\begin{array}[t]{@{}l@{}}
  [0.5,4.5],\ [5.5,9.5],\\
  {[10.5,14.5]},\ [15.5,19.5]~\mathrm{m}
  \end{array}\) \\
  Minimum PA spacing \(d_l^{\min}\) & \(0.1~\mathrm{m}\) \\
  Nominal feed coupling & \(\rho_m^{\rm F}=\rho_m^{\rm B}=1/\sqrt{2}\) \\
  Nominal PA extraction \(\zeta_n\) & \(0.5\) \\
  Extraction bounds \(\zeta_{\min},\zeta_{\max}\) & \(0.05/0.95\) \\
  Carrier frequency \(f_c\) & \(28~\mathrm{GHz}\) \\
  Propagation speed \(c_0\) & \(3\times10^8~\mathrm{m/s}\) \\
  Effective index \(n_{\rm eff}\) & \(1.45\) \\
  Guided attenuation \(\alpha_{\rm wg}\) &
  \(0.01842~\mathrm{m}^{-1}\) (\(0.08~\mathrm{dB/m}\)) \\
  Directional bandwidth \(B^a\) & \(10~\mathrm{MHz}\) \\
  UL power limit \(P_{r,\max}^{\rm u}\) & \(23~\mathrm{dBm}\) \\
  DL power budget \(P_{\max}^{\rm d}\) & \(10~\mathrm{W}\) \\
  Noise spectral density \(N_0\) & \(-174~\mathrm{dBm/Hz}\) \\
  Noise factors \(F_{\rm BS},F_r\) & \(10^{8/10}\) (\(8~\mathrm{dB}\)) \\
  Frame duration \(T\) & \(300~\mu\mathrm{s}\) \\
  Control \(\tau^{\rm u},\tau^{\rm e},\tau^{\rm d}\) & \(20/20/10~\mu\mathrm{s}\) \\
  Data-only \(\tau_0^{\rm u},\tau_0^{\rm d}\) & \(135/135~\mu\mathrm{s}\) \\
  Control bits \(L_i^{\rm u},L_i^{\rm d}\) & \(144/96\) bits \\
  Default control quality \(Q_{\rm ctrl}\) & \(30~\mathrm{dB}\) \\
  \bottomrule
 \end{tabular}
\end{table}

Each sampled double-integrator plant uses \(\mathbf Q_i=\diag(10,1)\),
\(\mathbf R_i=1\), and disturbance covariance
\(\mathbf W_i=\diag(10^{-4},10^{-3})\).
The default moderate gain \(\mathbf K_i\) is obtained by nominal
pole placement at \(0.998^{T/T_{\rm ref}}\) and \(0.995^{T/T_{\rm ref}}\),
where \(T_{\rm ref}=1~\mathrm{ms}\). The slow and fast gains use nominal
pole pairs \((0.9995^{T/T_{\rm ref}},0.9985^{T/T_{\rm ref}})\) and
\((0.9950^{T/T_{\rm ref}},0.9880^{T/T_{\rm ref}})\), respectively.
Stability is verified for the delayed full-update system. The control requirement is
\(\overline J_i=(1+\overline\Delta_{\rm ctrl})J_i(1)\), with
\(Q_{\rm ctrl}=-10\log_{10}\overline\Delta_{\rm ctrl}\), so the
default \(Q_{\rm ctrl}=30\) dB permits a \(0.1\%\) excess cost.
The corresponding reliability thresholds \(p_i^{\rm req}\) are
estimated by grid search and root refinement, with stability and cost
checked for each design.
Performance is measured by the control cost \(J_i(p_i)\)
in~\eqref{eq:general-control-cost}, the frame-average throughput \(C\)
in~\eqref{eq:communication-utility}, and the directional coexistence rate \(\widetilde C_{\rm co}^{a}
=B^a\sum_{j\in\mathcal K}\log_2(1+\gamma_j^{\mathrm{com},a})\)
and its duration-weighted aggregate
\(C_{\rm co}=(\tau^{\rm u}\widetilde C_{\rm co}^{\rm u}
+\tau^{\rm d}\widetilde C_{\rm co}^{\rm d})/
(\tau^{\rm u}+\tau^{\rm d})\).
Here, \(C_{\rm co}\) averages over the concurrent intervals, while
\(C\) also includes data-only access over the entire frame. Unless stated otherwise, throughput
curves report means over \(20\) paired layouts in Mbit/s.

\subsection{Control-Cost Analysis}
\label{sec:numerical-mapping}

Figure~\ref{fig:control-mapping}(a) shows close agreement between the
analytical control cost in~\eqref{eq:general-control-cost} and Monte Carlo
estimates for the three tested feedback gains under delayed actuation
and command holding. For each gain, the cost rises sharply as the
round-trip update probability decreases toward its stability threshold.
The sufficient reliability targets in panel (b) increase with
\(Q_{\rm ctrl}\), as less excess cost is allowed, and are higher for faster
feedback among the tested gains. Panel (c) maps these targets to
directional SINR thresholds using an equal UL/DL reliability split and
a common reliability-target schedule across actuation offsets. Shorter
control intervals require higher SINRs, reducing the admissible data
interference at a given desired-signal power
in~\eqref{eq:inner-performance-region}.

\subsection{Algorithm Performance}
We assess AO through throughput convergence and solver comparisons
under a common CPU-time budget.

\subsubsection{Convergence Analysis}
\label{sec:numerical-convergence}
\label{sec:numerical-reconfiguration}
\label{sec:numerical-spatial-load}
\begin{figure}[tb]
 \centering
 \includegraphics[width=0.85\columnwidth]{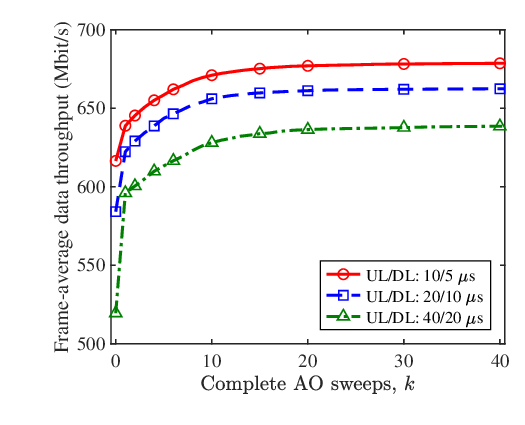}
 \vspace{-5mm}
 \caption{Throughput convergence of the proposed AO algorithm.}
 \label{fig:ao-convergence}
 \label{fig:pa-reconfiguration}
 \label{fig:spatial-load}
\end{figure}

Figure~\ref{fig:ao-convergence} plots frame-average throughput \(C\)
against complete AO sweeps under three control-transmission schedules.
Throughput increases rapidly and approaches a stable level after
approximately 10 AO sweeps. These results show the throughput convergence
of the proposed AO algorithm.

\subsubsection{Throughput Comparison}
\label{sec:numerical-algorithm-comparison}
\begin{figure}[tb]
 \centering
 \includegraphics[width=0.85\columnwidth]{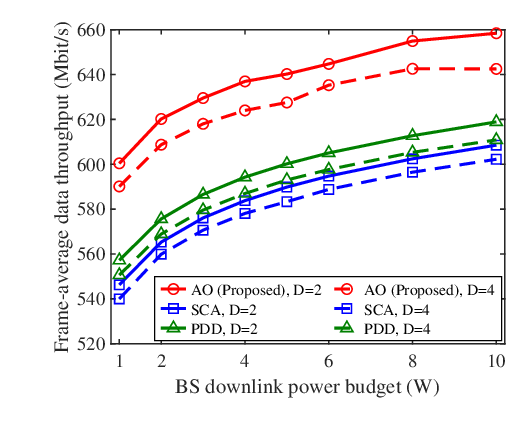}
 \vspace{-5mm}
 \caption{Frame-average throughput versus BS downlink power.}
 \label{fig:algorithm-comparison}
\end{figure}

Figure~\ref{fig:algorithm-comparison} compares AO with joint successive
convex approximation (SCA)~\cite{scutari2017constrainedsca} and penalty
dual decomposition (PDD)~\cite{shi2020pdd} under different BS powers
and control loads. Each method optimizes the same physical, four-stage
digital, and reliability variables from common feasible initial resources,
with 20 CPU seconds per power point. Under this budget, AO achieves
the highest mean throughput across the tested powers. Higher control loads
reduce throughput. The reason is that protecting more control streams
leaves fewer resources for data transmission.
\label{sec:numerical-independent-validation}

\subsection{Throughput Comparison of C-PASS and PASS}
\label{sec:numerical-boundary}
\begin{figure}[!t]
 \centering
 \includegraphics[trim=-33.3bp 0bp 25.2bp 0bp,clip,width=0.85\columnwidth]{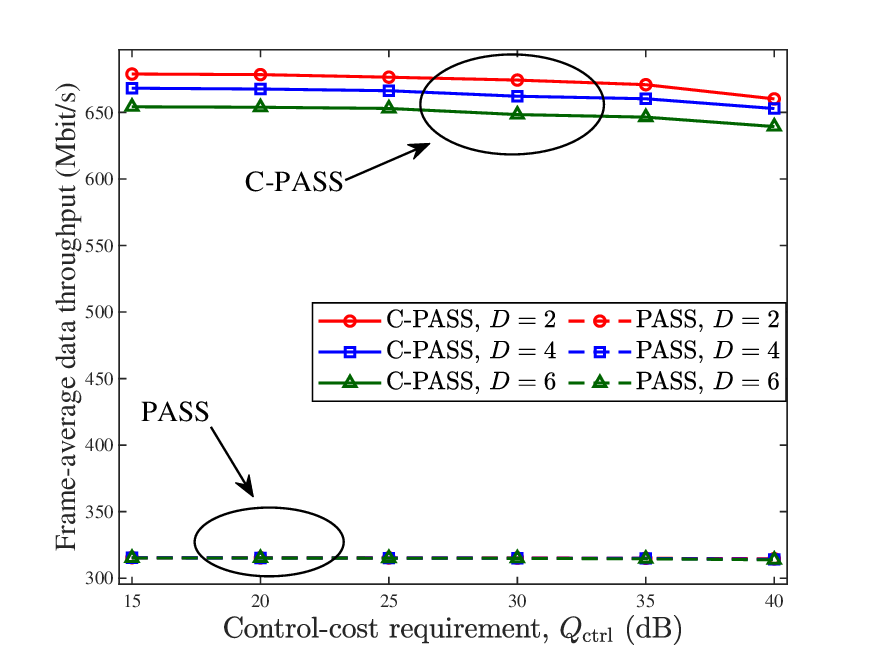}
 \vspace{-5mm}
 \caption{Frame-average throughput of C-PASS and PASS under different
 control-cost requirements and control loads.}
 \label{fig:performance-region}
\par
 \centering
 \includegraphics[width=0.85\columnwidth]{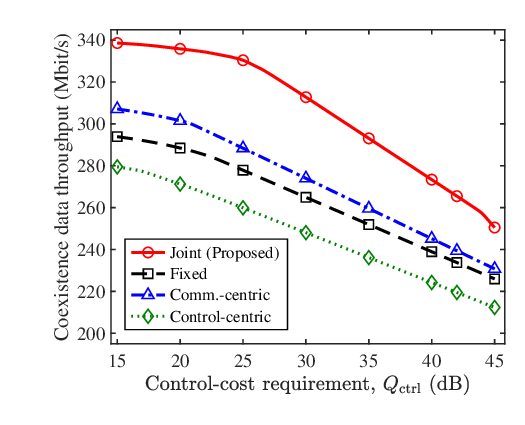}
 \vspace{-5mm}
 \caption{Coexistence throughput versus control-cost requirement.}
 \label{fig:cpass-boundary-extension}
\end{figure}

\begin{figure}[t]
 \centering
 \includegraphics[width=0.85\columnwidth]{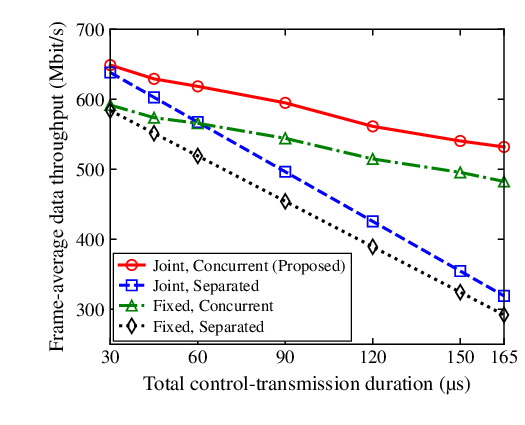}
 \vspace{-5mm}
 \caption{Frame-average throughput versus total control-transmission duration.}
 \label{fig:joint-service-load}
\end{figure}

Figure~\ref{fig:performance-region} compares the frame-average
throughput of C-PASS and end-fed PASS for \(D\in\{2,4,6\}\).
Both use two waveguides and eight PAs with identical placement
intervals under the same power and airtime budgets. C-PASS uses four
RF chains and admits concurrent data and control transmission,
whereas PASS uses two RF chains, serves control devices in pairs,
and restricts data transmission to data-only intervals. The additional
RF inputs and concurrent data access thus enable C-PASS to achieve
higher throughput throughout the tested settings. For example, at
\(Q_{\rm ctrl}=30\) dB and \(D=4\), C-PASS achieves \(662.16\) Mbit/s,
compared with \(315.04\) Mbit/s for PASS, yielding a \(110.2\%\) gain.
As \(Q_{\rm ctrl}\) or \(D\) increases, stronger protection of control
streams leaves fewer resources for data, reducing C-PASS throughput.
PASS throughput varies only slightly in these tests, as its data-only
airtime remains unchanged.

\subsection{Throughput Gains from Joint C-PASS Design}
\label{sec:numerical-cpass-extension}
Figure~\ref{fig:cpass-boundary-extension} compares four hardware policies
using \(C_{\rm co}\) with joint UL/DL design. \emph{Fixed} uses nominal
hardware. \emph{Communication-centric} and \emph{Control-centric} keep
their separately trained hardware fixed. The former excludes control
traffic during training, while the latter minimizes duration-weighted
normalized control-transmit energy at \(40\) dB with data muted.
\emph{Joint} adapts PA positions, feed coupling, and shared extraction
to maximize \(C_{\rm co}\) for each requirement, achieving the highest
throughput across the tested requirements. The gain over Fixed reflects
hardware adaptation that shapes desired signals and stream separation,
while the gain over Communication-centric shows the benefit of accounting
for control protection. The lower
Control-centric throughput further illustrates the cost of prioritizing
control energy over data service.

\subsection{Benefits of Concurrent Data and Control Transmission}
\label{sec:numerical-service-load}

Figure~\ref{fig:joint-service-load} compares Joint/Fixed hardware under
Concurrent/Separated transmission for \(D=7\). Concurrent admits data
streams in the control slots, while Separated mutes them. Joint adapts
one hardware configuration across all four stages. The UL/DL
duration ratio, frame duration, and computation time retain their
defaults, with remaining airtime split equally between data-only UL
and DL. The absolute control-cost cap remains fixed, while reliability
targets track the actuation delay. Under these settings, Joint--Concurrent
achieves the highest frame-average throughput, with a growing advantage
as control airtime increases. Longer control intervals provide more
channel uses but increase actuation delay and reduce data-only airtime.
Concurrent transmission sustains data service during these intervals,
allowing Joint--Concurrent to outperform Joint--Separated by \(66.7\%\)
at \(165~\mu\mathrm{s}\). The Joint--Separated/Fixed--Concurrent crossover
shows that control-slot reuse can outweigh hardware adaptation alone,
while the Joint--Fixed gap under Separated transmission shows gains
in the data-only channels. These results show that concurrent
transmission mitigates the throughput loss from longer control airtime,
with additional gains from hardware adaptation.

\section{Conclusion}\label{sec:conclusion}
This article developed an analytical and optimization framework for C-PASS communication
under prescribed control-cost limits. The analysis derived expressions for data throughput and control cost
under mutual interference during state uploading and command delivery. A finite-blocklength normal approximation converted
these limits into sufficient UL and DL SINR constraints. These constraints
defined an inner communication--control performance region and quantified
admissible data interference. The framework then formulated a throughput
maximization problem for joint C-PASS configuration and transceiver design.
An alternating algorithm combined sequential quadratic programming for
transceiver optimization with quasi-Newton updates for C-PASS reconfiguration.
Numerical results showed that the proposed C-PASS design achieved 110.2\%
higher frame-average throughput than the considered conventional PASS
baseline. By reusing control-transmission intervals for data service,
concurrent data and control transmission increased frame-average throughput
by up to 66.7\% over separated transmission.

\appendices

\section{Proof of Proposition~\ref{prop:reciprocal-feeding}}
\label{app:reciprocity-proof}
\label{app:reciprocal-control}
For the linear, time-invariant dielectric C-PASS with symmetric permittivity and permeability
tensors, the electric and magnetic phasor fields
\(\boldsymbol{\mathcal E}^a\) and \(\boldsymbol{\mathcal H}^a\),
\(a\in\mathcal A\), satisfy the Lorentz identity~\cite{svendsen2013reciprocity}
\[
 \oint_{\partial V}
 (\boldsymbol{\mathcal E}^{\rm d}\times\boldsymbol{\mathcal H}^{\rm u}
 -\boldsymbol{\mathcal E}^{\rm u}\times\boldsymbol{\mathcal H}^{\rm d})
 \cdot\mathrm d\mathbf A=0,
\]
where \(V\) is the propagation volume excluding the excitation
sources, and \(\mathrm d\mathbf A\) is the outward surface element
on its boundary \(\partial V\). With outgoing radiation at infinity,
this identity yields a reciprocity relation between transmitted and
received port signals. Expanding the boundary fields in normalized
port waves and selecting unit excitation at BS feed port \(m\) in
DL and at terminal \(r\) in UL, with other incident waves zero and
unchanged terminations, gives
\[
\begin{aligned}
 0=\sum_p\bigl(a_p^{\rm d}b_p^{\rm u}
              -a_p^{\rm u}b_p^{\rm d}\bigr)=[\mathbf h_r^{\rm u}]_m-[\mathbf h_r^{\rm d}]_m^*,
\end{aligned}
\]
where \(a_p^a\) and \(b_p^a\) are the incident and outgoing
power-normalized amplitudes at calibrated BS feed or terminal antenna
port \(p\), with real reference impedances.
The two surviving terms are the received UL and DL amplitudes, with
conjugation arising from the DL row convention
\((\mathbf h_r^{\rm d})^{\Hh}\). Since this equality holds for every
feed port \(m\), the channel vectors obtained
from~\eqref{eq:cpass-channel} are complex conjugates, proving the
reciprocity stated in Proposition~\ref{prop:reciprocal-feeding}.\hfill\(\square\)

\section{Proof of Mean-Square Stability and Control Cost}\label{app:sampled-control-model}
\label{subsec:sampled-control-model}
\label{app:control-cost-proof}
\label{app:control-cost}
\label{subsec:control-cost-proof}
Let \(\mathcal H_{i,k}\) collect the history, augmented state, and current
update outcome before the disturbance. The conditional moments
\(\E[\mathbf w_{i,k}\mid\mathcal H_{i,k}]=\zero\) and
\(\E[\mathbf w_{i,k}\mathbf w_{i,k}^{\T}\mid\mathcal H_{i,k}]=\mathbf W_i\)
eliminate state--disturbance cross terms. Averaging the delivery outcomes
then gives~\eqref{eq:covariance-recursion}
and~\eqref{eq:expected-stage-cost}. Without process disturbances,
\[
\begin{aligned}
 \vecop(\boldsymbol\Sigma_{i,k})
 &=\mathbf M_i(p)^k\vecop(\boldsymbol\Sigma_{i,0}),\\
 \E[\|\mathbf z_{i,k}\|^2]&=\tr(\boldsymbol\Sigma_{i,k}).
\end{aligned}
\]
The standard second-moment stability
criterion~\cite{costa1993stability,schenato2009hold} therefore gives
\(\E[\|\mathbf z_{i,k}\|^2]\to0\) for every finite initial second moment
if and only if \(\rho(\mathbf M_i(p))<1\), proving
\eqref{eq:mean-square-stability}. At \(p=1\),
\(\mathbf M_i(1)=\mathbf F_{i,1}\otimes\mathbf F_{i,1}\), so the
prescribed stabilizing gain gives
\(\rho(\mathbf M_i(1))=\rho(\mathbf F_{i,1})^2<1\).
With process disturbances and \(\rho(\mathbf M_i(p))<1\), the initial-state
contribution vanishes, while the accumulated disturbance converges through
\(\sum_{t=0}^{\infty}\mathbf M_i(p)^t=[\I-\mathbf M_i(p)]^{-1}\).
This gives the unique stationary second
moment~\eqref{eq:general-closed-form-covariance}. Its time average has the
same limit, so substitution into~\eqref{eq:control-cost-moment-average}
gives~\eqref{eq:general-control-cost}. This completes the proof of
Remark~\ref{rem:control-cost-convergence}.\hfill\(\square\)

\bibliographystyle{IEEEtran}
\bibliography{reference/jdcc}
\end{document}